\documentclass[prd,onecolumn,floatfix]{revtex4}
\usepackage{amssymb}
\usepackage{epsfig}
\usepackage{amsmath}
\usepackage{verbatim}

\input{tcilatex}

\begin{document}

\title{Vacuum structure and string tension in Yang-Mills dimeron ensembles}
\author{Falk Zimmermann$^{1,2}$, Hilmar Forkel$^{1}$ and Michael M\"{u}%
ller-Preu\ss ker$^{1}$}
\affiliation{$^{1}$Institut f\"{u}r Physik, Humboldt-Universit\"{a}t zu Berlin, D-12489
Berlin, Germany}
\affiliation{$^{2}$HISKP and Bethe Center for Theoretical Physics, Universit\"{a}t Bonn,
D-53115 Bonn, Germany}

\begin{abstract}
We numerically simulate ensembles of SU$\left( 2\right) $ Yang-Mills dimeron
solutions with a statistical weight determined by the classical action and
perform a comprehensive analysis of their properties as a function of the
bare coupling. In particular, we examine the extent to which these ensembles
and their classical gauge interactions capture topological and confinement
properties of the Yang-Mills vacuum. This also allows us to put the classic
picture of meron-induced quark confinement, with the
confinement-deconfinement transition triggered by dimeron dissociation, to
stringent tests. In the first part of our analysis we study spacial,
topological-charge and color correlations at the level of both the dimerons
and their meron constituents. At small to moderate couplings, the dependence
of the interactions between the dimerons on their relative color
orientations is found to generate a strong attraction (repulsion) between
nearest neighbors of opposite (equal) topological charge. Hence the emerging
short- to mid-range order in the gauge-field configurations screens
topological charges. With increasing coupling this order weakens rapidly,
however, in part because the dimerons gradually dissociate into their less
localized meron constituents. Monitoring confinement properties by
evaluating Wilson-loop expectation values, we find the growing disorder due
to the long-range tails of these progressively liberated merons to generate
a finite and (with the coupling) increasing string tension. The
short-distance behavior of the static quark-antiquark potential, on the
other hand, is dominated by small, \textquotedblleft
instanton-like\textquotedblright\ dimerons. String tension, action density
and topological susceptibility of the dimeron ensembles in the physical
coupling region turn out to be of the order of standard values. Hence the
above results demonstrate without reliance on weak-coupling or low-density
approximations that the dissociating dimeron component in the Yang-Mills
vacuum can indeed produce a meron-populated confining phase. The density of
coexisting, hardly dissociated and thus instanton-like dimerons seems to
remain large enough, on the other hand, to reproduce much of the additional
phenomenology successfully accounted for by non-confining instanton vacuum
models. Hence dimeron ensembles should provide an efficient basis for a
rather complete description of the Yang-Mills\ vacuum.
\end{abstract}

\maketitle
\preprint{HU-EP-12/06}
\issuenumber{HU-EP-12/06}

\section{Introduction}

\label{introsec}

The notorious complexity of the Yang-Mills vacuum manifests itself in
emergent infrared phenomena which, despite vigorous theoretical efforts over
the last decades, remain only fragmentarily understood. Quark confinement 
\cite{mil} was early recognized as a paradigm for these infrared
complexities, both because of its unprecedented nature and of its almost
universal impact on the hadronic world. The long struggle to explain the
nonperturbative confinement mechanism has led to the development and often
ongoing refinement of a wide variety of theoretical approaches \cite{conftec}%
, ranging from various vacuum models to ab-initio lattice simulations \cite%
{wil74etc}. The latter, in particular, have provided strong and unbiased
evidence for the main signatures of quark confinement, i.e. the linear
growth (and saturation) of the static quark-antiquark potential at large
distances and the associated color flux-tube formation \cite{ftl} (and
breaking \cite{bal05}).

Beyond establishing confinement and its characteristics, intense efforts
were devoted to pinning down the underlying dynamical mechanism and to
identify the infrared degrees of freedom best suited to describe it in
humanly fathomable terms. Vacuum\ model analyses have played a prominent and
often pioneering role in this endeavor. The majority of them is based on
ensembles of specific \textquotedblleft constituent\textquotedblright\
fields, i.e. at least partly localized\ gauge fields (with their prospective
monopole loop and center vortex content) which are supposed to populate and
disorder the Yang-Mills vacuum. Potential candidates for these building
blocks can be classified according to the increasing number of spacetime
dimensions in which they are localized, and thus equivalently to their
increasing efficiency in disordering the vacuum. The minimally, i.e. in just
two dimensions localized candidates are center vortices \cite{tho78,vort}.
An intermediate position occupy both (gauge-projected) Abelian \cite%
{che97,tho76} and non-Abelian monopoles or dyons \cite{susyqcd,man04}
(possibly in the guise of BPS constituents \cite{dia09} of KvBLL calorons 
\cite{kra98}, see below) which are localized in the three spacial dimensions.

Among the maximally, i.e. in all four (Euclidean) spacetime dimensions
localized candidates for the building blocks of confining field
configurations, finally, are regular-gauge instantons \cite{len08}, calorons
with nontrivial \cite{kra98,bru07,cal-latt,ger07} holonomy and merons \cite%
{dea76,len04,len08}. All of these \textquotedblleft
pseudoparticles\textquotedblright\ carry a nontrivial topological charge,
mediate vacuum tunneling events and solve the classical Yang-Mills equation 
\footnote{%
One may abandon the requirement that the constituent fields are Yang-Mills
solutions, e.g. by multiplying them with variable coefficients \cite{wag05}.
The latter approach leads to a more flexible vacuum description but prevents
a semiclasssical interpretation.}. The latter property, in particular, has
raised hopes that at least one of these solutions may provide a basis for
semiclassical and therefore potentially analytical treatments of
confinement. (This proved indeed possible in lower-dimensional models where
instantons or vortices were shown to generate confinement at weak coupling 
\cite{pol77,kog01}.) In classically scale-invariant, four-dimensional
Yang-Mills theory, however, where one is inevitably faced with large
couplings at large distances, the justification for semiclassical
approximations depends crucially on the characteristic distance scales
involved. A successful semiclassical description of confinement, in
particular, would require all physically relevant confinement effects (as,
for example, the quark binding inside light hadrons) to take place over
distances which remain small enough to\ avoid the strong-coupling regime.

In order to overcome these potentially stymying limitations, one may resort
to simulating fully interacting pseudoparticle ensembles numerically. The
first and still most extensive such calculations were based on rather
dilute\ superpositions of instantons and anti-instantons in singular gauge 
\cite{sch98}. Such superpositions maintain part of the instantons'
semiclassical nature, are known to generate a variety of\ important physical
effects including spontaneous chiral symmetry breaking, and describe a large
amount of successful vacuum and hadron phenomenology \cite{sch98}. However,
they fail to generate a confining, i.e. linearly growing potential between
static quarks \cite{cal78,cal378,dia89}. By now, numerical simulations were
performed for superpositions of all the other above-mentioned pseudoparticle
candidates as well. A central result was that each of them proved capable of
generating a finite string tension.

The discovery of new finite-temperature instantons\ \cite{kra98} has given
renewed impetus to semiclassically inspired confinement models. These
generalized calorons contain $N_{c}$ self-dual monopole (\textquotedblleft
dyon\textquotedblright ) constituents without classical interactions and
carry a non-trivial holonomy as an explicit degree of freedom. The ability
of the calorons to separate into $N_{c}$ quasi-independent constituents
(without cost in action) in the confinement phase has led to simulations 
\cite{ger07} where the separation appears as an additional (internal) degree
of freedom compared to the \textquotedblleft
old-fashioned\textquotedblright\ instanton simulations. Encouraging results
have been obtained. Later, an analytic solution for the caloron-dyon gas was
proposed \cite{dia09} at the price of a non-positive weight of an
overwhelming multitude of configurations with respect to the moduli metric 
\cite{bru09}. An analytic as well as numerical treatment of the purely
Abelian non-interacting dyon gas was shown to provide a confining force \cite%
{bru11}.

In addition, equilibrated lattice gauge-field configurations were
successfully searched for the presence of pseudoparticles (with the notable
exception of merons) in the Yang-Mills vacuum. Such searches are typically
performed by smoothing techniques designed to filter out infrared gluon
fields and in particular solutions of the classical Yang-Mills equation. For
examples in the most extensively studied case of instantons and their size
distribution see Refs. \cite{instlat}, and for calorons with nontrivial
holonomy Refs. \cite{cal-latt}.

In the present paper, building upon the above developments, we are going to
introduce and study a new type of pseudoparticle ensemble. The latter is
based on superpositions of dimerons and anti-dimerons (in singular gauge),
i.e. on those four-dimensionally localized solutions of the classical SU$%
\left( 2\right) $ Yang-Mills equation which contain two meron centers \cite%
{dea76}. The relatively large dimeron solution family includes as one
extreme members which are contracted to (singular-gauge) instantons and as
another members which are dissociated into two far separated merons.
Moreover, it contains continuous sets of dimeron fields which interpolate
between these extremes. Hence our configuration space includes many of the
previously studied singular-gauge instanton\ and single-meron configurations
and is ideally suited to study transitions between instanton and meron
ensembles. This is a crucial advantage because such a transition process was
conjectured to take place between the instanton, dimeron and meron
components of the Yang-Mills vacuum as a function of the bare gauge
coupling. Indeed, this transition is a central ingredient of the
meron-induced confinement scenario which was put forward on the basis of
qualitative arguments by Callan, Dashen and Gross (CDG) \cite{cal277,cal78}
(and further explored in several promising directions, including e.g. a
two-phase picture of hadron structure \cite{cal279} which recieved some
support from lattice simulations \cite{fab95}). (For more recent work on
merons see Refs. \cite{newmeron}. An alternative meron confinement mechanism
was suggested in Refs. \cite{gli78}.)

The CDG confinement scenario was inspired by the already mentioned
weak-coupling confinement mechanism in 2+1 dimensional Yang-Mills-Higgs
theory \cite{pol77,kog01}. In contrast to this classic paradigm, however,
even the ingenious use of semiclassical arguments turned out to be of
limited use in 3+1 dimensional Yang-Mills theory. In fact, the resulting
estimates remained qualitative at best \footnote{%
In the words of Sidney Coleman \cite{col80}: \textquotedblleft I can see
nothing wrong with this idea in principle, but the details of the argument
involve a stupendous amount of hand-waving. This part is just a suggestion
(although a very clever suggestion) that may or may not someday become a
theory of confinement.\textquotedblright\ }. Further progress was achieved
only recently, when a numerical study of superpositions of merons and
antimerons \cite{len04,len08} provided the first convincing support for a
central assumption of the CDG scenario, namely that the meron component of
the Yang-Mills vacuum may indeed confine, i.e. generate a static quark
potential which rises linearly at large interquark separations. The
suggested origin of the meron population in the vacuum from dimeron
dissociation remained obscure, on the other hand, since the field content
was from the outset restricted to a purely meronic phase.

To search for the\ development of such a meron-dominated phase and to
analyze the underlying mechanism in dimeron-antidimeron ensembles evolving
under the Yang-Mills dynamics will therefore be important points on our
agenda. For this purpose, and in contrast to previous simulations of
pseudoparticle ensembles, we will monitor pertinent vacuum properties in a
wide range of gauge coupling values. In particular, we will look for
signatures of the envisioned dimeron dissociation process. Similar to the
gradual break-up of molecules into their atomic constituents with increasing
temperature, this dissociation is suggested to be driven by a competition
between the decreasing attraction among the dimeron's regularized meron
centers and the entropy \footnote{%
The dimeron's entropy is the logarithm of its collective-coordinate (or
moduli) space volume.} gain from their increasingly distant positions.
Guided by the behavior of an ideal dimeron gas, one may indeed roughly
estimate the inter-meron attraction to decrease relative to the entropy with
growing coupling. In view of the strong interactions anticipated especially
between the meron centers of different dimerons in significantly dissociated
ensembles, however, such estimates are unreliable. Moreover, in our
dynamical context the coupling-dependent competition between energy and
entropy is particularly subtle since both attraction and entropy are
expected to depend only logarithmically on the intermeron separation 
\footnote{%
As in several lower-dimensional systems, in the thermodynamic limit the
dimeron dissociation process may therefore lead to a sharp transition of
Kosterlitz-Thouless type \cite{kos73,cal77} from a deconfining to a
confining phase.}. This further impedes even qualitative analytical
estimates of dimeron ensemble properties.

Our fully interacting ensembles, on the other hand, are well suited to
tackle and clarify these issues. Since they take all classical gauge
interactions -- including those which may be strong, long-range or many-body
-- between the dimerons into account, we do not have to rely on the
semiclassical or any other weak-coupling or low-density approximation.
Nevertheless, it will be of heuristic value and of help for future, possibly
more analytic treatments to understand which aspects of the dimeron-ensemble
dynamics may be described semiclassically. In fact, one of our reasons for
transforming the individual dimerons into a singular gauge was to provide
additional insight into this issue by more strongly localizing them. This
allows the dimerons and their meron centers to retain their identities and
their classical shapes to a larger extent, even in the presence of quantum
fluctuations, and thus facilitates (semi-)classical behavior. At several
points during the course of our investigation we will therefore check for
indications of such behavior.

Besides studying the dimeron dissociation process and its dynamics in
detail, we will also survey other structural and topological\ properties of
the ensemble configurations. In particular, we will examine distributions
and correlations of the topological charge carriers and search for ordering
tendencies with respect to both dimeron and meron centers. (A side benefit
for investigating topology distributions is that dimerons, in contrast to
e.g. center vortices and monopoles \cite{rei02}, carry the topological
charge of the Yang-Mills gauge group \footnote{%
i.e. the 2nd Chern class of the associated principal fiber bundle}
directly.) This will result in a more detailed understanding of the
restructuring processes which accompany the transition to the
strong-coupling regime. We will pay particular attention to the behavior of
the topological susceptibility and of confinement properties, which we
monitor by evaluating Wilson-loop expectation values and the associated
string tension, as a function of the bare coupling. The quantitative picture
emerging from these investigations will reveal, in particular, at which
stage of dissociation (as quantified e.g. by the average separation of the
dimerons' meron partners) dimeron ensembles can best describe the confining
phase of the Yang-Mills vacuum.

In the next section we start by acquainting the reader with pertinent
properties of the dimeron solutions, transform them into a singular gauge
and discuss the regularization of their singularities. We then set up our
ensemble field configurations and write down their partition function.
Section \ref{simul} presents an outline of our simulation strategy and, in
particular, of the measures taken to control systematic finite-size and
discretization errors. The next and central Sec. \ref{ressec} contains our
results and their discussion. We start with an extensive statistical
analysis of the spacetime, topological-charge and color structure of the
ensemble configurations as well as of the crucial dimeron dissociation
process. These investigations provide a broad spectrum of insights into the
behavior of the dimeron ensemble and its phase structure as a function of
the gauge coupling. They furthermore relate this behavior to the changes in
the average properties of the individual dimerons. We then evaluate the
topological susceptibility and the static quark potential (based on the
calculation of Wilson-loop expectation values) and analyze the results in
relation to the coupling-dependent restructuring of the ensemble
configurations. We further discuss scale-setting issues, evaluate
dimensionless observable ratios, rewrite our results in physical units and
confront them with those of other approaches. Section \ref{sum}, finally,
contains a summary of our main findings and presents our conclusions.

\section{Singular-gauge dimeron configurations and their dynamics}

\label{sdeft}

In the following subsections we construct the (continuum) dimeron
field-configuration space on which our study will be based. Along the way,
we will motivate our expectation that such dimeron configurations
approximate crucial features of the gauge-field population in the Yang-Mills
vacuum, and in particular that they may provide a decent description of the
confinement-deconfinement transition with decreasing gauge coupling. We
start by deriving the building blocks of our fields, i.e. the individual
dimeron solutions of the Yang-Mills equation in singular gauge, and discuss
their collective coordinates. We then superpose these solutions to define
our gauge-field configuration space and write down the partition function
which specifies its dynamics. Finally, we discuss the (both gauge-dependent
and -independent) singularities of our configurations and introduce suitable
regularization procedures to prepare for their numerical treatment in Sec. %
\ref{simul}.

\subsection{Yang-Mills dimerons in singular gauge}

\label{sgmeronsec}

Dimerons (or meron pairs) are those classical solutions of the Euclidean
Yang-Mills equation which contain two meron centers where their only
singularities are located \cite{dea76}. In regular gauge, with the two
merons located symmetrically at the distances $\pm a$ from the center at $%
x_{0}$, the SU$\left( 2\right) $ dimeron solution family reads \cite{dea76}%
\begin{equation}
A_{\mu }^{\left( \text{D},\text{r}\right) }\left( x;\left\{
x_{0},a,u\right\} \right) =\left[ \frac{\left( x-x_{0}+a\right) _{\nu }}{%
\left( x-x_{0}+a\right) ^{2}}+\frac{\left( x-x_{0}-a\right) _{\nu }}{\left(
x-x_{0}-a\right) ^{2}}\right] u^{\dagger }\sigma _{\mu \nu }u  \label{dimreg}
\end{equation}%
where $\sigma _{\mu \nu }:=\eta _{a\mu \nu }\sigma _{a}/2$ with the Pauli
matrices $\sigma _{a}$ and the 't Hooft symbols $\eta _{a\mu \nu }$. (The
latter are defined as $\eta _{a\mu \nu }=\varepsilon _{a\mu \nu }$, $\eta
_{a\mu 4}=-\eta _{a4\mu }=\delta _{a\mu }$ for $\mu ,\nu =1,2,3$ and $\eta
_{a44}=0$ \cite{tho276}.) The unitary matrices $u$ are global SU$\left(
2\right) $ color rotations. The dimerons (\ref{dimreg}) carry the same
topological charge as instantons, i.e. $Q=1$. The anti-dimeron $\bar{A}_{\mu
}^{\left( \text{D},\text{r}\right) }$ with opposite topological charge $Q=-1$
is obtained from Eq. (\ref{dimreg}) by replacing $\eta _{a\mu \nu
}\rightarrow \bar{\eta}_{a\mu \nu }:=\left( -1\right) ^{\delta _{\mu
4}+\delta _{\nu 4}}\eta _{a\mu \nu }$. In contrast to instantons, however,
dimerons are not selfdual.

The solution class (\ref{dimreg}) depends on eleven real and continuous
\textquotedblleft collective coordinates\textquotedblright\ or
\textquotedblleft moduli\textquotedblright\ whose values uniquely and
completely specify each member. Eight of these parameters, $x_{0,\mu }$ and $%
a_{\mu }$, determine the spacetime position and orientation of the dimeron
while the remaining three determine the SU$\left( 2\right) $ group elements $%
u$ \footnote{%
More generally, these are the $N_{c}^{2}-1$ \textquotedblleft Euler
angles\textquotedblright\ which parametrize (constant) SU$\left(
N_{c}\right) $ group elements.}. The collective coordinates correspond to
those eleven independent combinations of the classical and continuous
Yang-Mills symmetries, i.e. of Euclidean spacetime translations and
rotations, conformal transformations and global SU$\left( 2\right) $ color
rotations, which transform a representative dimeron into a
gauge-inequivalent solution \cite{gid79}.

It is instructive to consider several limits of the solution family (\ref%
{dimreg}). For $a\rightarrow 0$ the meron centers coalesce and the dimeron
turns into a pointlike, regular-gauge instanton. (A finite-size instanton is
reached after suitably regularizing the meron singularities, cf. Sec. \ref%
{reg}.) For $\left\vert a\right\vert \rightarrow \infty $, on the other
hand, the dimeron breaks up into two merons. Although these merons maintain
the fixed relative color orientation of Eq. (\ref{dimreg}), changing it by
hand will yield increasingly action-degenerate two-meron solutions when the
inter-meron separation $\left\vert a\right\vert $ becomes large. This is
because the color-dependent attraction which locks the meron constituents of
Eq. (\ref{dimreg}) into their rigid color orientation decreases with
increasing $\left\vert a\right\vert $. Hence both instantons and merons are
contained in the solution class (\ref{dimreg}) as limiting cases. The fields
(\ref{dimreg}) therefore provide on-shell interpolations between instantons
and isolated meron pairs, i.e. continuous paths in Yang-Mills solution space
which connect these particular gauge fields. In our context these paths are
of particular interest since they provide preferred doorways along which the
dimerons may dissociate into merons. As already mentioned, such dissociation
processes are conjectured to drive the deconfinement-confinement transition.

The solutions (\ref{dimreg}) have the same $\left\vert x\right\vert
\rightarrow \infty $ behavior as an instanton in regular gauge \cite%
{bel75,col80,vai99}, i.e. they contain a long-distance tail 
\begin{equation}
A_{\mu }^{\left( \text{D},\text{r}\right) }\left( x\right) \overset{%
\left\vert x\right\vert \rightarrow \infty }{\longrightarrow }2\frac{x_{\rho
}}{x^{2}}\sigma _{\mu \rho }  \label{regdimtail}
\end{equation}
$\sim 1/\left\vert x\right\vert $ which implies an exceptionally weak
localization. In multi-dimeron configurations these tails generate strong
overlap interactions between the individual dimerons. Building dimeron
ensembles by superposing solutions of the type (\ref{dimreg}) would thus
lead to very strongly correlated systems, in some respects similar to the
regular-gauge instanton ensembles studied in Ref. \cite{len08}. In the
present work we will take a different route, however. In contrast to merons,
dimerons -- like instantons -- may be transformed to singular gauges in
which they are more strongly localized, with their long-range tails decaying
as $1/\left\vert x\right\vert ^{3}$. A superposition of singular-gauge
dimerons thus improves the vacuum description at short distances, compared
to superpositions of either regular-gauge instantons or single merons. As
long as the characteristic dimeron size is small compared to the average
inter-dimeron separation, furthermore, it will provide much better
approximations to classical Yang-Mills solutions. This is essentially
because at moderate pseudoparticle densities the contribution of the overlap
regions to the action is much smaller than in regular-gauge dimeron
superpositions. Hence singular-gauge multi-dimeron configurations provide a
privileged testing ground especially for the semiclassical features of the
CDG confinement mechanism \cite{cal277,cal78,cal279}.

In addition, dimeron configurations in singular gauge may also describe less
classical or even fully quantum-mechanical aspects of confinement. This
holds, in particular, for the vacuum disordering mechanism envisioned by
CDG. The latter relies on the long-distance tails of the individual merons
which come into play when the separation $\left\vert a\right\vert $ between
the meron centers of the dimerons becomes large. As discussed above, the
partner merons are then practically independent and regain their long-range
tail $\sim 1/\left\vert x\right\vert $ (at least in isolation). The ensuing
long-distance color correlations among these merons were found in Refs. \cite%
{len04,len08} to be far from semi-classical and to generate confinement. Of
course, a complete dimeron dissociation into isolated and thus far separated
merons is impossible in the finite spacetime volumes in which numerical
simulations are feasible. (The same holds for the solution in which one of
the merons delocalizes on the three-sphere at spacetime infinity \cite%
{dea76,cal78}.) However, such isolated merons would anyhow be unphysical
(since they carry infinite action) and incompatible with a finite dimeron
density. Instead, one expects that beyond some typical separation the two
meron centers of a dimeron experience stronger interactions with their
individual field environment than with their partner. The rigid link between
their color orientations can then be broken, i.e. the partner merons can
become effectively independent of each other.

In order to construct the multi-dimeron configurations motivated above, we
first transform the dimeron solution family (\ref{dimreg}) into a specific
singular\ gauge,%
\begin{equation}
A_{\mu }^{\left( \text{D},\text{s}\right) }=\hat{U}^{\dagger }\left( \hat{x}%
\right) A_{\mu }^{\left( \text{D},\text{r}\right) }\hat{U}\left( \hat{x}%
\right) +i\hat{U}^{\dagger }\left( \hat{x}\right) \partial _{\mu }\hat{U}%
\left( \hat{x}\right) ,
\end{equation}%
where $\hat{U}$ is the large, i.e. topologically active gauge group element 
\begin{equation}
\hat{U}\left( \hat{x}\right) =i\hat{x}_{\mu }\sigma _{\mu }^{\left( +\right)
}  \label{sgtrf}
\end{equation}%
($\sigma _{\mu }^{\left( +\right) }\equiv \left( \vec{\sigma},-i\right) ,$ $%
\hat{x}_{\mu }\equiv x_{\mu }/\sqrt{x_{\nu }x_{\nu }}$) which has a
singularity at the origin of the coordinate system. The result can be cast
into the form%
\begin{equation}
A_{\mu }^{\left( \text{D},\text{s}\right) }=f_{1}\left( x,a\right) \text{ }%
x_{\nu }\bar{\sigma}_{\mu \nu }+f_{2}\left( x,a\right) \text{ }a_{\nu
}X_{\mu \nu }\left( \hat{x}\right)  \label{gmsing}
\end{equation}%
($\bar{\sigma}_{\mu \nu }:=\bar{\eta}_{a\mu \nu }\sigma _{a}/2$). Choosing
for simplicity $x_{0}=0$ and $u=1$, the two scalar functions $f_{1,2}$
become 
\begin{align}
f_{1}\left( x,a\right) & :=\frac{2}{x^{2}}-\frac{1}{\left( x+a\right) ^{2}}-%
\frac{1}{\left( x-a\right) ^{2}},  \label{f1} \\
f_{2}\left( x,a\right) & :=\frac{1}{\left( x+a\right) ^{2}}-\frac{1}{\left(
x-a\right) ^{2}}.  \label{f2}
\end{align}%
The antisymmetric and anti-selfdual field $X_{\mu \nu }$ contains important
parts of the SU$\left( 2\right) $ and spacetime tensor structures (and their
mutual couplings). Explicitly, 
\begin{equation}
X_{\mu \nu }\left( \hat{x}\right) =\eta _{a\mu \nu }X_{a}\left( \hat{x}%
\right) \text{ \ \ \ with }X_{a}\left( \hat{x}\right) =\left( \frac{1}{2}-%
\hat{x}_{r}\hat{x}_{r}\right) \sigma _{a}+\hat{x}_{a}\hat{x}_{s}\sigma _{s}-%
\hat{x}_{4}\hat{x}_{s}\varepsilon _{asc}\sigma _{c}
\end{equation}%
(latin indices are spacial). After regularization of the short-distance
singularity (cf. Sec. \ref{reg}) the $a\rightarrow 0$ limit turns the
dimeron (\ref{gmsing}) into a singular-gauge instanton with $f_{1}\left(
x,a=0\right) =2\rho ^{2}/\left[ x^{2}\left( x^{2}+\rho ^{2}\right) \right] $
and $f_{2}\left( x,a=0\right) =0$. In the following, we will often refer
with the term \textquotedblleft dimeron\textquotedblright\ to both dimerons
and anti-dimerons in singular gauge. Occasionally, we will use the term
\textquotedblleft pseudoparticle\textquotedblright\ for the same purpose.

The leading asymptotic behavior of the singular-gauge dimerons (\ref{gmsing}%
) is%
\begin{equation}
A_{\mu }^{\left( \text{D},\text{s}\right) }\overset{\left\vert x\right\vert
\gg \left\vert a\right\vert }{\longrightarrow }-4\frac{a_{\rho }x_{\rho }}{%
x^{4}}a_{\nu }\eta _{a\mu \nu }X_{a}\left( \hat{x}\right) \sim \frac{1}{x^{3}%
},
\end{equation}%
i.e. the long-range tail (\ref{regdimtail}) has as intended disappeared and
the overlap between neighboring dimerons is strongly reduced. (This is in
contrast to the meron ensembles of Refs. \cite{len04,len08} whose
constituents exist only in regular gauge.) In addition to the meron-center
singularities at $x=\pm a$, the solution (\ref{gmsing}) inherits another
singularity at $x=0$ from the gauge transformation (\ref{sgtrf}). Hence the
impact of the latter will disappear when forming (topologically insensitive)
gauge-invariant quantities from Eq. (\ref{gmsing}). The regularization of
these singularities will be discussed in Sec. \ref{reg}.

By construction, the singular-gauge dimerons (\ref{gmsing}) provide only a
subset of the full solution family. The complete eleven-parameter family can
be recovered by translating the solutions (\ref{gmsing}) to $x_{0}$ and by
gauge-rotating them with a constant matrix $u\in $ SU$\left( 2\right) $, i.e.%
\begin{equation}
A_{\mu }\left( x;\left\{ x_{0},a,u\right\} \right) =u^{\dagger }A_{\mu
}^{\left( \text{D},\text{s}\right) }\left( x-x_{0}\right) u=R_{ab}\left(
u\right) A_{\mu }^{\left( \text{D},\text{s}\right) ,b}\left( x-x_{0}\right) 
\frac{\sigma _{a}}{2}
\end{equation}%
where the orthogonal matrices $R_{ab}\left( u\right) =tr\left\{ \sigma
_{a}u^{\dagger }\sigma _{b}u\right\} /2$ are obtained from the condition $%
u^{\dagger }\sigma _{b}u=R_{ab}\left( u\right) \sigma _{a}$. The group
elements $u$ depend on three real and continuous parameters which form
(local)\ coordinates on the SU$\left( 2\right) $ group manifold $S^{3}$. In
the following it will be convenient to use the quaternion representation%
\begin{equation}
u\left( c\right) =c_{0}+ic_{a}\sigma _{a}  \label{quat}
\end{equation}%
which embeds $S^{3}$ into $R^{4}$ with Euclidean coordinates $%
c_{0},...,c_{3} $ by imposing the unit three-sphere constraint $\sum_{\alpha
=0}^{3}c_{\alpha }^{2}=1$. In these coordinates $R$ takes the form 
\begin{equation}
R_{ab}\left( u\left( c\right) \right) =\left( c_{0}^{2}-c_{c}c_{c}\right)
\delta _{ab}+2c_{a}c_{b}-2\varepsilon _{abc}c_{0}c_{c}.
\end{equation}

For our discussion below it will be useful to keep in mind that dimerons
have three more continuous and noncompact collective coordinates than
instantons (but three less than a pair of independent merons, due to the
locking of the inter-meron color orientation). The three additional
coordinates arise from the more complex structure of extended (i.e. $%
\left\vert a\right\vert \neq 0$) dimerons which requires more degrees of
freedom to locate and orient them in spacetime.\ This results in a
substantially larger \textquotedblleft position entropy\textquotedblright\ 
\cite{cal78,cal278} which is instrumental in counterbalancing the
(regularized) dimeron\ action which grows logarithmically with $\left\vert
a\right\vert $. Indeed, the larger entropy is a necessary (but not
sufficient) requirement for dimerons to dissociate with increasing coupling $%
g^{2}$ and to finally split up into their meron partners.

\subsection{Field configurations and partition function of the dimeron
ensembles}

\label{fc}

As motivated above, we intend to study a model which drastically reduces the
field content of Yang-Mills theory (as integrated over in amplitudes and the
partition function) to superpositions of $N_{D}$ dimerons and $N_{\bar{D}%
}\simeq $ $N_{D}$ antidimerons in singular gauge \footnote{%
It may be useful to recall that the superposition of singular-gauge dimerons
is \emph{not} gauge-equivalent to a superposition of regular-gauge dimerons.}%
, i.e. 
\begin{equation}
A_{\mu }\left( x,\left\{ C_{I,i},\bar{C}_{\bar{I},i}\right\} \right)
=\sum_{I}^{N_{D}}A_{\mu }^{\left( \text{D},\text{s}\right) }\left( x;\left\{
C_{I,i}\right\} \right) +\sum_{\bar{I}}^{N_{\bar{D}}}\bar{A}_{\mu }^{\left( 
\text{D},\text{s}\right) }\left( x;\left\{ \bar{C}_{\bar{I},i}\right\}
\right)  \label{cf}
\end{equation}%
with $N\equiv $ $N_{D}+N_{\bar{D}}$. Each term in this sum is uniquely
characterized by the set $\left\{ C_{I}\right\} =\left\{
x_{0,I},a_{I},c_{I}\right\} $ of collective coordinates of the corresponding
(anti-) dimeron. We recall that the configurations (\ref{cf}) differ
distinctly from those obtained by transforming a regular-gauge dimeron
superposition into (any) singular gauge. This is because in Eq. (\ref{cf})
the gauge of each dimeron is chosen relative to its individual position. One
may wonder, incidentally, whether $n$-meron solutions with $Q>1$ and their
anti-solutions should be added to the superposition ansatz (\ref{cf}). As in
the case of instantons, however, the Bogomoln'yi-type bound $S\geq \left(
8\pi ^{2}/g^{2}\right) \left\vert Q\right\vert $ on the action \cite{bel75}
implies that such multi-(anti)-meron contributions to the partition function
are exponentially suppressed relative to the dimeron contributions \footnote{%
From the practical point of view this is fortunate since no multi-meron
solutions with $\left\vert Q\right\vert >1$ seem to be known analytically.}.
Since the entropy increases only logarithmically with $n$ and thus cannot
compensate this suppression, such multi-meron contributions may be safely
neglected.

Nevertheless, the ansatz (\ref{cf}) should be regarded as a rather minimal
choice. It is mainly geared towards a transparent study of the proposed
(instanton and) dimeron dissociation mechanism and its role in the
deconfinement-confinement transition \cite{cal277,cal78,cal279}. Hence there
are several natural directions in which Eq. (\ref{cf}) may be extended in
future studies to provide a more complete description of the Yang-Mills
vacuum physics. An example would be to add the meron-antimeron pair
solutions \cite{dea76} of the Yang-Mills equations, again individually
transformed into singular gauge. This would maintain the approximately
semiclassical nature of the configurations at small $\left\vert a\right\vert 
$ and yield a richer dynamics. However, it would probably also lead to a
less transparent interpretation of the results, and we do not expect the
topologically trivial meron-antimeron pair configurations to provide
qualitatively new insights into the transition behavior. Indeed, their
limits for $\left\vert a\right\vert \rightarrow 0$, a pure-gauge field of
zero action, and for $\left\vert a\right\vert \rightarrow \infty $, a meron
and an independent antimeron, indicate that meron-antimeron pairs do not
generate new pathways for the transition. More promising improvement options
would include generalizations of the dimeron superposition ansatz (\ref{cf})
which allow for a complete transition into a meron ensemble (e.g. by
releasing the rigid color locking between the dimerons' meron partners
beyond a suitable intermeron separation $\left\vert a\right\vert _{\min }$)
or the admixture of a pure instanton component \cite{cal78,cal279} with a
realistic size distribution \cite{rin99}.

As discussed above, we view the dimeron configurations (\ref{cf}) as a
pertinent subset of the SU$\left( 2\right) $ gauge fields governed by the
Yang-Mills dynamics. Hence we define the partition function of our dimeron
model as%
\begin{equation}
Z=\int \dprod\limits_{I,\bar{I}}^{N_{D},N_{\bar{D}}}dC_{I}d\bar{C}_{\bar{I}%
}\exp \left\{ -S\left[ A_{\mu }\left( \left\{ C_{I,i},\bar{C}_{\bar{I}%
,j}\right\} \right) \right] \right\}  \label{pf}
\end{equation}%
where $S$ is the Euclidean Yang-Mills action%
\begin{equation}
S\left[ A\right] =\frac{1}{2g^{2}}\int d^{4}xtr\left\{ F_{\mu \nu }F_{\mu
\nu }\right\} =:\int d^{4}xs\left( x\right)  \label{sym}
\end{equation}%
and $F_{\mu \nu }$ is the gauge field strength 
\begin{equation}
F_{\mu \nu }\left( x\right) =\partial _{\mu }A_{\nu }\left( x\right)
-\partial _{\nu }A_{\mu }\left( x\right) -i\left[ A_{\mu }\left( x\right)
,A_{\nu }\left( x\right) \right] .  \label{fym}
\end{equation}%
As already indicated, we expect the gauge interactions among the dimerons in
Eq. (\ref{pf}) to play an important role in generating confining long-range
correlations for large $\left\vert a\right\vert $ and, in particular, to
provide a disordering mechanism for the vacuum. This is in contrast to the
situation in non-confining singular-gauge instanton ensembles where for most
amplitudes a random orientation of the instantons (i.e. the neglect of the
interaction term $\exp \left( -S\right) $ in Eq. (\ref{pf})) yields a
reasonable approximation to the vacuum physics \cite{sch98}.

The integration over the collective coordinates will be performed with the
measure%
\begin{equation}
dC=d^{4}x_{0}d^{4}ad\mu \left( \vec{c}\right)
\end{equation}%
($d\mu \left( \vec{c}\right) $ is the SU$\left( 2\right) $ Haar measure) for
the dimerons, and analogously for the anti-dimerons. This type of measure if
familiar from instanton vacuum models \cite{sch98}. An improved alternative
may include Jacobians which arise from the transformation of the linear
gauge-field measure $dA$ into the collective-coordinate basis and thereby
implement the full moduli-space metric. Finally, an additional $a$
dependence of the measure can emerge from the trace anomaly \cite%
{cal78,cal279}.

\subsection{Regularization of the gauge-field singularities}

\label{reg}

The singular-gauge dimeron solutions (\ref{gmsing}) contain two types of
singularities. The first are those located at the constituent meron centers $%
x=x_{0}\pm a$ which persist in any gauge. In addition, there is a gauge
singularity inherited from the topologically large and thus necessarily
singular gauge transformation (\ref{sgtrf}), chosen to sit at the origin. In
isolated dimerons, such gauge singularities could be gauged away and would
therefore not affect (topologically insensitive) observables. This ceases to
be the case for the superpositions (\ref{cf}), however, where the location
of the gauge singularities varies with the positions of the pseudoparticles.

Since both types of singularities would impede our numerical simulations,
they have to be regularized in a physically acceptable manner. (Recall that
the individual dimeron fields (\ref{dimreg}) and (\ref{gmsing}) do not solve
the Yang-Mills equation at their singularities, so that a suitably localized
regulator will only minimally affect the semiclassical properties of the
configurations (\ref{cf}).) A natural way to regularize the singularities at
the meron centers is to add the square of a \textquotedblleft
size\textquotedblright\ parameter $\rho $ to the denominators $\left(
x-x_{0}\pm a\right) ^{2}$ in Eqs. (\ref{f1}) and (\ref{f2}). Such a
regulator imitates the way in which the size parameter enters the instanton
solutions. Hence it may arise from scale-symmetry breaking quantum
fluctuations which are expected to smear the classical singularities. A
finite $\rho $ furthermore acts as a UV cutoff since it limits the gradients
of the configurations (\ref{cf}). Of course, the regularized field
configurations cease to be exact solutions of the classical Yang-Mills
equation. Hence the value of $\rho $ should be chosen large enough to avoid
sizeable discretization errors (cf. Sec. \ref{eran}) but also small enough
to avoid unnessary deformations of the dimeron solutions from the
semiclassical saddle points. We have found $\rho ^{2}=0.3$\ to be a
reasonable compromise between both criteria and will use this value in our
numerical simulations, if not stated otherwise.

The remaining singularities are, at least for isolated dimerons, gauge
artefacts. They remain unlikely to have a physical impact in our dimeron
superpositions, too, but they may nevertheless cause problems in numerical
approximations (even for single dimerons), e.g. when subtle cancellations in
gauge-invariant quantities are upset by discretization errors. A finite-size
regulator as above would hardly help to avoid such problems since it would
just spread out the unphysical action-density peaks at the singularities.
Hence we treat these gauge-dependent singularities in a pragmatic way,
namely by interpolating the gauge field in 4$d$ balls of radius $\varepsilon 
$ around the singularities with the field value at a specific point on the
surface of the ball. In practice, it turns out that with $\rho ^{2}=0.3$
these highly localized singularities are smoothly regularized by taking $%
\varepsilon $ as small as $\varepsilon =5\times 10^{-10}$. (The values of
dimensionful quantities, such as $\rho $ and $\varepsilon $ above, are given
in \textquotedblleft numerical units\textquotedblright\ originating from the
discretization grid to be introduced in Sec. \ref{box}. As a convenient
length unit we have chosen $a=0.1$, i.e. a tenth of the distance between
nearest grid points. (Of course, this $a$ should not be confused with the
four-vector $a$ which parametrizes the dimeron solutions.) When transforming
our results into physical units starting from Sec. \ref{topdissec}, we will
denote quantities in these numerical units by a hat above their symbols, cf.
Sec. \ref{topdissec}.) In our simulations it typically took several hundred
sweeps through configurations of $N=487$ pseudoparticles (cf. Sec. \ref%
{mcsec}) before such a singularity was first encountered. (Even with a
dramatically reduced floating-point variable length of 16 bits,
incidentally,\ we have observed no overflows (which unregularized gauge
singularities would generate) when evaluating the action density of a
regularized two-dimeron configuration.)

\section{Simulation details}

\label{simul}

In the following subsections we summarize how we have generated the dimeron
configuration ensembles on which all our subsequent calculations will be
based. We further discuss predominant sources of systematic errors in these
ensembles.

\subsection{Sampling volume and resolution}

\label{box}

Owing to the translational invariance of the Yang-Mills dynamics (\ref{sym}%
), ensemble averages over the pseudoparticle configurations (\ref{cf}) in
infinite spacetime will likewise be translationally invariant. In our
simulations we are restricted to a bounded volume of numerically manageable
size, on the other hand, and thus have to keep boundary artefacts in
calculated amplitudes under control, ideally within the size of the
statistical uncertainties. Our initial step in this direction was to adopt 
\emph{singular-gauge} dimerons with their more rapidly decaying long-range
tails as the constituents of our gauge-field configurations (\ref{cf}). A
sufficient suppression of boundary artefacts turned out to require
additional measures, however, since the overlap among the still rather
moderately localized dimerons generate important (and typically repulsive)
interactions over ranges beyond the average nearest-neighbor distance. In
fact, these interactions are expected to play an important role in the
confinement mechanism and should thus be distorted as little as possible by
the boundary. Simulation costs, on the other hand, with their by far largest
part due to integrating the action density on a spacetime grid, should be
kept minimal.

\begin{figure}[tbp]
\begin{center}
\includegraphics[height = 5cm]{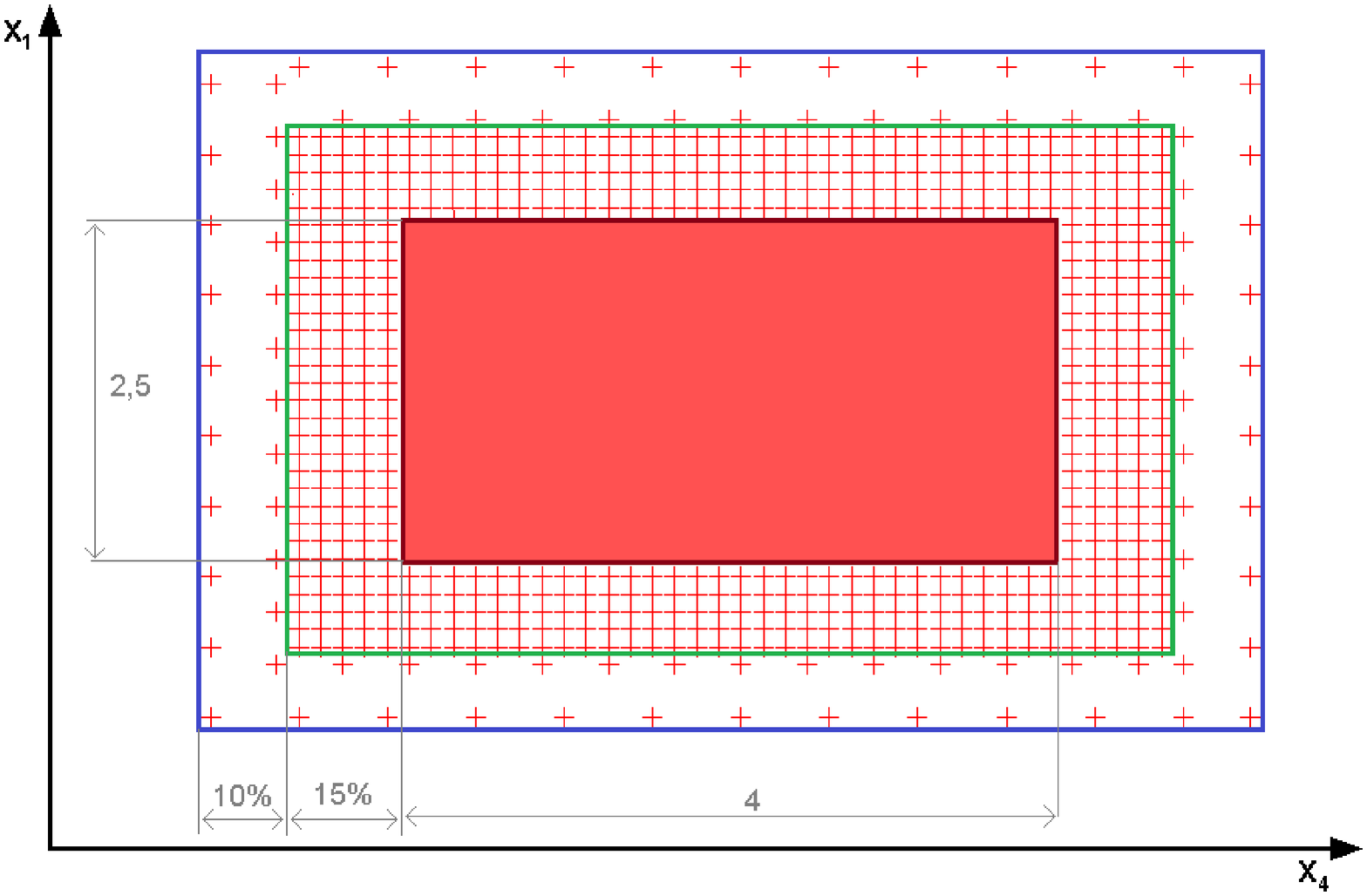}
\end{center}
\caption{The multi-layered multigrid designed to control boundary effects
when numerically evaluating the action density.}
\label{gridfig}
\end{figure}

In order to approximately meet these conflicting goals, we will adopt a
multi-layered description of the boundary with decreasing grid-point density
towards the outer layers, as sketched in Fig. \ref{gridfig}. The innermost
\textquotedblleft core\ volume\textquotedblright\ is a rectangular spacetime
box in which we intend to evaluate amplitudes and thus require the physics
to best approximate the infinite-volume limit. In order to allow for an
efficient evaluation of Wilson loops with elongated time directions in Sec. %
\ref{wlsec}, this volume and its grid are chosen to be asymmetric. One
dimension, singled out as the Euclidean time direction, contains 40 grid
points while each spacial dimension contains 25. This turns out to provide a
sufficiently high resolution for the amplitudes to be calculated below (cf.
Sec. \ref{eran}).

The core volume is encompassed by an \textquotedblleft ensemble\
volume\textquotedblright\ which contains the (anti-) meron centers of all
(anti-) dimerons. Its surface is implemented numerically by rejecting
Metropolis updates (cf. Sec. \ref{mcsec}) during which a meron center would
leave this volume. The extent of the ensemble volume in a given direction
will be taken 15\% larger than that of the core in the same direction. In
the part of the ensemble box which surrounds the core we reduce the
grid-point density to $\left( 2/3\right) ^{4}\sim 20\%$ of that in the core
volume. This turns out to yield a still adequate resolution while
substantially reducing the computational cost of evaluating the action.

A satisfactory suppression of field distortions inside the core volume turns
out to require an additional precautionary measure, however. It consists in
correcting for a particularly prominent boundary effect, namely the
artificial attraction\ of the pseudoparticles to the surface of the ensemble
volume. The latter arises because a substantial part of the action density
of dimerons near the boundary is located outside the ensemble volume and
thus not accounted for while the compensating tails of outside dimerons
reaching into the ensemble volume are neglected. We approximately remove
this artefact by surrounding the ensemble box with another,
\textquotedblleft covering\textquotedblright\ volume which extends beyond
the ensemble box by 10\% of the core size in each direction. To keep the
additional computational costs under control, we decrease the grid-point
density of the outermost shell (i.e. the part of the covering volume not
shared by the ensemble volume) inversely with the distance from the ensemble
boundary. Since this shell does not contain the rapidly varying fields close
to the meron centers, calculating the action in the full covering volume
indeed largely prevents the fake attraction to the ensemble boundary (cf.
Sec. \ref{dimenssec}). Alternatively, finite-size effects in pseudoparticle
ensembles can be efficiently corrected by employing the Ewald summation
technique, as has been shown very recently in the simpler case of Abelian
dyon field ensembles \cite{bru11}.

\subsection{Monte-Carlo updates with dynamical resolution and step-size
adaptation}

\label{mcsec}

We evaluate the discretized functional integral over the pseudoparticle
fields, i.e. the multi-dimensional integral over their collective
coordinates in the partition function (\ref{pf}) or any other amplitude,
stochastically by Monte-Carlo importance sampling. Hence we average over
dimeron configurations chosen randomly from a Gibbs distribution with
Boltzmann factor $\exp \left( -S\right) $ where $S$ is the Yang-Mills action
(\ref{sym}). More specifically, we use the Metropolis algorithm to generate
homogeneous Markov sequences of dimeron configurations that visit fields
with larger probabilities more often. After reaching equilibrium, the
probability of finding a configuration in the ensemble of subsequently
generated fields follows the Gibbs distribution.

The initial pseudoparticle configurations for these Markov sequences are
obtained by choosing their collective coordinates from a uniform random
distribution. This procedure has equal access to all members of the
configuration space and practically always results in configurations far
from equilibrium, with actions several orders of magnitude above their
equilibrium values. We have also tested various more ordered initial
arrangements of the dimerons and convinced ourselves that those lead within
errors to equivalent equilibrium ensembles (see Sec. \ref{eran}).

For the Markov update of the $n$-th configuration $A_{n}$ to its sequel $%
A_{n+1}$ according to the Metropolis rules, we first generate some candidate
configuration $A^{\prime }$ by randomly choosing an individual
pseudoparticle in $A_{n}$ and modifying alternately (from one candidate to
the next) either its position, i.e. $a_{\mu }\rightarrow a_{\mu }+da_{\mu }$
and $x_{0,\mu }\rightarrow x_{0,\mu }+dx_{0,\mu }$, or its color orientation 
$c_{i}\rightarrow c_{i}+d\mu \left( c_{i}\right) $. The increments $da,$ $%
dx_{0}$ and $dc$ are chosen randomly and uniformly\ subject to the
constraint that their length remains limited. More specifically, we demand $%
\left\vert dx_{0}\right\vert \in \left[ 0,\left\vert dx_{0}\right\vert _{%
\text{max}}\right] $, $\left\vert da\right\vert \in \left[ 0,\left\vert
da\right\vert _{\text{max}}\right] $ as well as $\left\vert dc\right\vert
\in \left[ 0,\left\vert dc\right\vert _{\text{max}}\right] $ \footnote{
More precisely, all four coefficients of the quaternion representation 
(\ref{quat})\ are varied independently. The constraint among them is 
reinstated only afterwards by rescaling all coefficients with a common 
factor. 
(Hence $d {\protect \mu} \left( c_{i} \right)$ represents the Haar measure 
on SU$\left(2\right)
\sim S^{3}$.)} and further restrict $\left\vert da\right\vert _{\text{max}%
}=\left\vert dx_{0}\right\vert _{\text{max}}$. The candidate configuration $%
A^{\prime }$ is accepted as the new configuration $A_{n+1}$ with probability 
$\min \left\{ 1,\exp \left( S\left[ A\right] -S\left[ A^{\prime }\right]
\right) \right\} $. If a candidate is rejected, the unchanged configuration
is taken as $A_{n+1}$ (and included in measurements like all others).

The choice of the maximal modification step sizes $\left\vert da\right\vert
_{\text{max}}$ and $\left\vert dc\right\vert _{\text{max}}$ can be used\ to
optimize both the thermalization rate and the decorrelation among subsequent
field configurations of the generated ensembles. This requires a compromise
between too small values, which impede the progression through configuration
space, and too large values which more strongly decorrelate subsequent
configurations but cause an unefficiently large update rejection rate.
Examination of the relation between maximal step sizes and typical
acceptance probabilities suggests an optimal acceptance rate of about 25\%
for updates in both position and color space. During the initial
thermalization process we will therefore increase (decrease) the maximal
step sizes $\left\vert da\right\vert _{\text{max}}$ and $\left\vert
dc\right\vert _{\text{max}}$ after every 10 consecutive candidate field
configurations by 10\% if the average acceptance rate for these
configurations falls below (rises above) 25\%.

This dynamical step-size adaptation procedure accelerates the approach to
equilibrium and avoids getting trapped into approximate would-be equilibria.
Initially, i.e. far from equilibrium and near the high-action random
configurations, the fields\ can relax in larger steps while closer to
equilibrium the action fluctuations become smaller and require decreasing
step sizes to maintain sufficient acceptance rates. From the time when
approximate equilibration sets in, however, the step size is kept constant
to preserve detailed balance. We further note that increasing $g$ increases
the acceptance probability since the Yang-Mills action (\ref{sym}) scales as 
$g^{-2}$. Hence larger $g$ allow for larger step sizes, accelerate
thermalization (in less Markov steps) by decreasing the autocorrelation time
and result in equilibrium ensembles with larger entropy.

\begin{figure}[tbp]
\begin{center}
\includegraphics[height = 12cm]{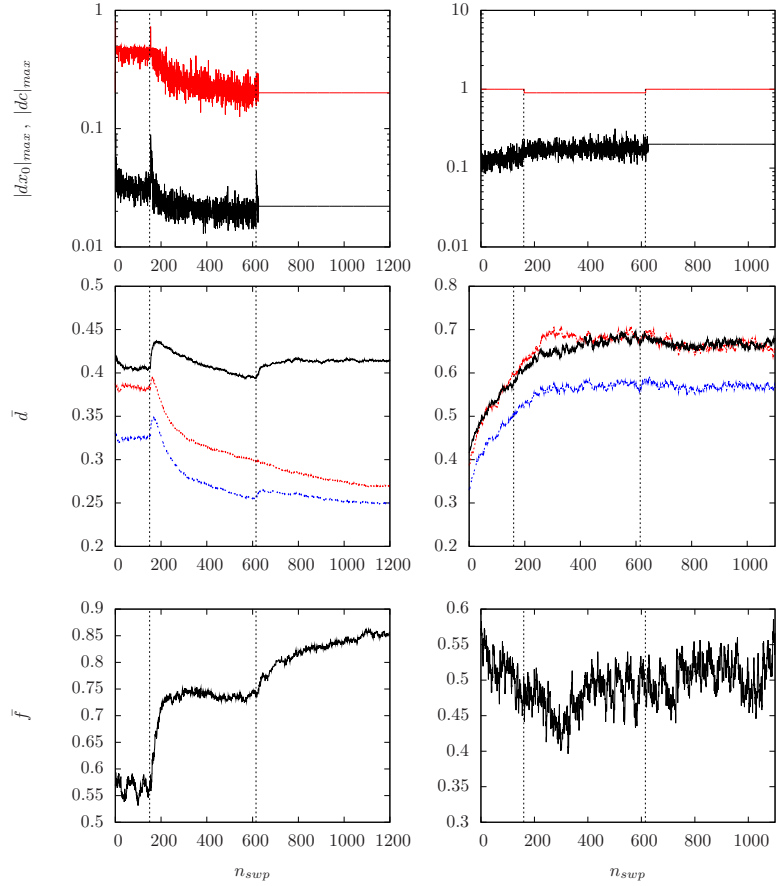}
\end{center}
\caption{The thermalization histories of simulations with $g^{2}=1$ (left
panels) and $g^{2}=25$ (right panels) as a function of the number $n_{swp}$
of sweeps. The uppermost row\ shows the evolution of the maximal step widths
in position (lower, black curves) and color (uper, red curves)\ space. In
the second row we plot the average distances $\bar{d}$ between
nearest-neighbor dimerons of equal (black curves, uppermost for $g^{2}=1$),
opposite (red curves, intermediate for $g^{2}=1$) and arbitrary
(lowest-lying, blue curves) topological charge. The last row shows the
average probability $\bar{f}$ for the nearest neighbor of a given meron
center to have opposite topological charge. (The dotted vertical lines
indicate the sweep numbers at which the resolution of the action sampling is
increased. For a stringent test of the resolution quality we enhance the
corresponding action changes by smearing the meron-center singularities less
broadly, corresponding to $\protect\rho ^{2}=0.2$, in the $g^{2}=1$
simulation (left panels). For $g^{2}=25$ (right panels) we use our standard
value $\protect\rho ^{2}=0.3$.)}
\label{thermhistory}
\end{figure}

As a case in point, for $g^{2}=1$ (and $\rho ^{2}=0.3$) we find $\left\vert
da\right\vert _{\text{max}}\sim 0.05$ and $\left\vert dc\right\vert _{\text{%
max}}\sim 0.1$ in equilibrium, where the scale of the fluctuations is set by
the competition between action and entropy. (For $\rho ^{2}=0.2$, i.e. for
more strongly localized meron centers, one instead reads off the smaller
values $\left\vert da\right\vert _{\text{max}}\sim 0.025$ and $\left\vert
dc\right\vert _{\text{max}}\sim 0.2$ from Fig. \ref{thermhistory}.) For $%
g^{2}=10^{2}$, on the other hand, one finds the indeed substantially larger
maximal step sizes $\left\vert da\right\vert _{\text{max}}\sim \left\vert
dc\right\vert _{\text{max}}\sim 0.5$ (again for $\rho ^{2}=0.3$). The
decreasing step width during a typical thermalization history is plotted in
the uppermost row of Fig. \ref{thermhistory} for $g^{2}=1$ and $g^{2}=25$.
Note that different step sizes in position and color space are generally
required to obtain action changes of comparable magnitude.

The efficiency of the ensemble generation process can be further improved by
exploiting the during equilibration decreasing ruggedness and action of the
dimeron configurations in yet another way, namely by starting on coarser
grids and increasing the resolution only when a higher accuracy of the
action evaluation becomes necessary. Indeed, as long as the action density
is comparable to or larger than its value\ around the regularized meron
centers, a reduced resolution is generally sufficient and can save computer
time. During thermalization the overall action decreases strongly, however,
and the grid must be refined to prevent the then more prominent meron
centers from artificially reducing their action density by \textquotedblleft
hiding\textquotedblright\ between grid points. In practice, we implement
this refinement procedure by starting the simulations with a $\left(
1/3\right) ^{4}$ times smaller grid-point density. When the values of
suitable quantities (e.g. the distance between nearest-neighbor meron
centers) begin to saturate, this factor is increased to $\left( 2/3\right)
^{4}$. Only after again reaching approximate saturation the full grid-point
density of Sec. \ref{box} is activated for the final approach to equilibrium
and the subsequent generation of the thermal ensembles. The approximate
saturation plateaus and subsequent grid refinements (indicated by dotted
vertical lines)\ are clearly visible in the thermalization histories of Fig. %
\ref{thermhistory}.

In the following we will denote\ a set of $2N$ consecutive Markov steps as a
\textquotedblleft sweep\textquotedblright\ (where $N=N_{D}+N_{\overline{D}}$%
, cf. Sec. \ref{fc}). During such sweeps each pseudoparticle of a
configuration is on average once considered for a full update. (The factor
two arises since the individual Markov steps attempt to modify either a
dimeron's position or its color orientation, i.e. only one of the two
subsets of its degrees of freedom.) After four consecutive sweeps with the
acceptance rate kept fixed at $1/4$, all pseudoparticles in a configuration
are therefore on average updated once. In order to reduce autocorrelations
among ensemble configurations and thus to increase the statistical
independence of successive measurements, we will only select the
configurations generated by every fifth sweep (after approximate
equilibration of the Markov chain) as ensemble members and employ a binning
procedure to calculate amplitudes and their errors (cf. Sec. \ref{statsec}).

\subsection{Approach to equilibrium}

\label{equilsec}

As indicated in the previous section, a reliable calculation of vacuum
expectation values as ensemble averages requires a sufficient thermalization
of the Markov sequences which generate the ensemble configurations. Our
criteria for when their distribution approximates the equilibrium
distribution closely enough are based on the measurement of several
observables and correlations to be described below.

First insights into the configurations' thermalization properties can be
gained by following the Markov evolution of the average distance $\bar{d}$
of a fixed meron center to its nearest neighbor as a function of the number
of sweeps. Two examples for such histories are depicted in the second row of
Fig. \ref{thermhistory}. In the left panel the coupling is set to $g=1$,
i.e. our smallest value, which results in the slowest equilibration rates we
have to deal with in this paper (cf. Sec. \ref{mcsec}). (For these
measurements we have also selected a smaller meron-size regulator $\rho
^{2}=0.2$ which further slows the equilibration process.) A first\ and
general insight to be read off from this figure is how the degree of
thermalization reached after a given number of sweeps depends on the
measured quantity. While the average distance to the closest neighboring
meron of equal topological charge appears to have equilibrated after about
800 sweeps, the analogous distance for oppositely charged merons may not be
fully thermalized even after 1200 sweeps. A similarly long equilibration
history can be observed for the average probability $\bar{f}$ that the
nearest neighbor of a given meron center has opposite topological charge, as
plotted in the last row of Fig. \ref{thermhistory}. This behavior seems to
indicate that the average distance between two meron centers from the same
dimeron equilibrates faster than their average separation from the meron
centers of neighboring dimerons. For $g^{2}=25$ the typical relaxation times
reduce to at least half of those at $g^{2}=1$, confirming the expectation
that more strongly coupled ensembles thermalize faster.

For a better understanding of the global thermalization behavior, and
especially of the character and uniqueness of the reached equilibria, we
have also compared ensembles resulting from different initial
configurations. More specifically, we have prepared a set of initial dimeron
arrangements in which the pseudoparticles were regularly positioned at
maximal average distances. Their topological charges, color orientations and
sizes were chosen either to approximate different types of action minima or
to imitate close-to-equilibrium configurations found in previous
thermalization runs.

All ensembles resulting from these different initializations turned out to
generate within errors the same averages. This provides, for once, evidence
for the ergodicity of the underlying Markov process, i.e. for its ability to
access any possible dimeron configuration in a finite number of steps. More
importantly, this result strongly supports the conclusion that the
thermalization processes indeed have come sufficiently close to equilibrium,
i.e. that our measurements in Sec. \ref{ressec} are performed in almost
thermalized ensembles. This conclusion is strengthened by the fact that all
our runs with non-random initializations were performed at $g^{2}=1$ where
equilibration is particularly slow.

\subsection{Finite-size and discretization errors}

\label{eran}

We now turn to the analysis of the systematic errors which arise from
finite-resolution and finite-size effects. We then describe the steps taken
to control them by accordingly refining our calculational strategy.
Additional error-reduction measures, which apply to the calculation of
specific amplitudes only, will be outlined in their corresponding sections
below.

Discretization errors arise in our context from the practical necessity to
sample the field configurations (\ref{cf}) on spacetime grids of finite
resolution. To meet our accuracy requirements, we adapt the grid-point
density according to the characteristic length and gradient scales in the
pseudoparticle configurations under consideration. During thermalization,
where these scales change drastically, we do so dynamically as described in
Sec. \ref{mcsec}. In the (approximately)\ thermalized ensembles, on the
other hand, these scales are essentially fixed by the pseudoparticle density
and by the size of the regularized meron-center singularities \footnote{%
The singular-gauge induced singularities are so strongly localized (even
after regularization, cf. Sec. \ref{reg}), on the other hand, that they have
practically no impact on the simulations.}. As described in Sec. \ref{reg},
the regulator is chosen about five times larger than the lattice unit $a=0.1$%
, at $\rho \simeq \allowbreak 0.55$. This proves sufficient to keep the
action of an isolated dimeron practically independent\ of its position on
the grid. Although configurations with gradients larger than those around
single meron centers frequently occur among the dimeron superpositions (\ref%
{cf}), their enhanced action renders them relatively unimportant when
thermalization is achieved. This explains why we did not encounter such
configurations in our equilibrated ensembles (cf. Sec. \ref{dimenssec}).

The grid resolution's impact on the calculated action values and on the
thermalization process can be seen directly in the Markov evolution
histories of Fig. \ref{thermhistory}. The two vertical lines indicate the
sweep numbers at which the resolution of the grid is refined (cf. Sec. \ref%
{mcsec}). The full resolution, corresponding to $a=0.1$, is reached only
after the second refinement step. In order to provide a particularly
stringent test of discretization errors, we have reduced our regulator $\rho 
$ for this simulation by a factor of four, to $\rho =0.14$. This size is of
the order of the minimal lattice unit and about half of the value $\rho \sim
0.3$ below which discretization errors become a concern. As a consequence,
one observes that the evolution of both plotted quantities (i.e. the average
meron-center distances $\left\langle d_{\text{nearestMM}}\right\rangle $ and
the probability for the nearest neighbor-meron to have opposite $Q=\pm 1/2$)
on the coarsest grid begins to saturate at a preliminary would-be
equilibrium when further relaxation is prevented by insufficient
action-density resolution. Already after the first grid refinement, however,
relaxation continues until the deviation from the thermal values reduces to
about 15 -- 20\%. The equilibrium values are approached after the second
refinement step. For our standard meron size $\rho \simeq 0.55$ the
discretization errors will of course be much smaller and should be well
under control. This conclusion is confirmed by checks on other observables,
e.g. when calculating link elements and their concatenations into Wilson
loops in Sec. \ref{wlsec}.

We now turn to the discussion of finite-size effects. Those are artefacts of
the simulation volume's boundary and turn out to be more difficult to
control than the discretization errors. Our choice of the more strongly
localized dimerons in singular gauge (\ref{gmsing}) as the constituents of
the configurations (\ref{cf}) was partly motivated by reducing such boundary
effects. For the same purpose we designed the multi-layered boundary
outlined in Sec. \ref{box}. Nevertheless, sizeable boundary artefacts of
different origins, of different $g^{2}$ dependence and with varying impact
on the calculated quantities remain to be dealt with. In order to analyze
those quantitatively, we start by monitoring violations of translational
invariance in the ensemble-averaged action density $\left\langle s\left(
x\right) \right\rangle $ as a function of the distance from the boundary for
two values of $g^{2}$. The results are of direct physical and practical
importance since $\left\langle s\right\rangle $ is related to the gluon
condensate and since a reliable evaluation of the action is indispensable
for generating trustworthy ensembles.

\begin{figure}[tbp]
\begin{center}
\includegraphics[height = 7cm]{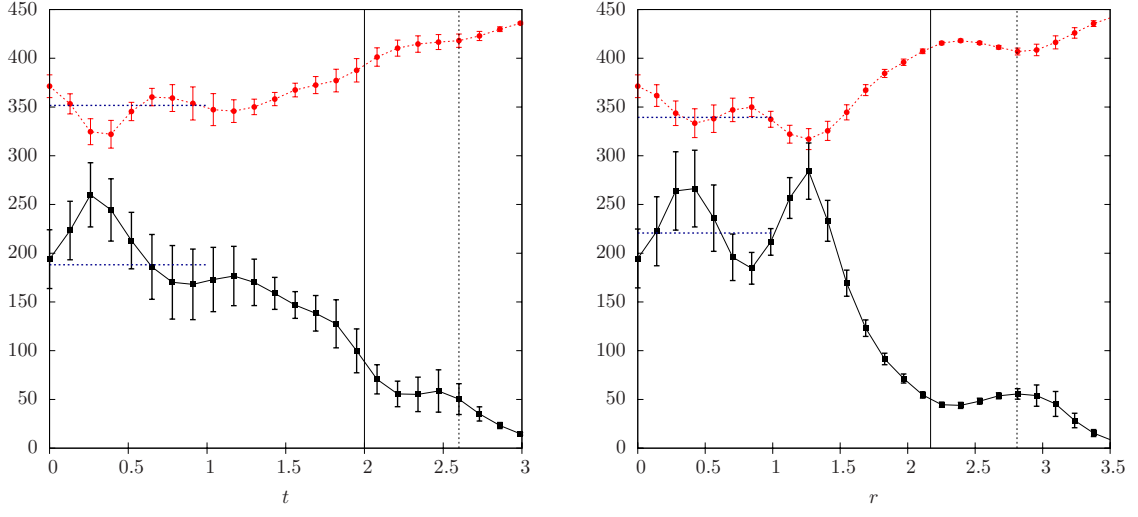}
\end{center}
\caption{The averaged action density $\left\langle s\right\rangle $ (black
squares, full line) and quadratic \textquotedblleft
plaquette\textquotedblright\ Wilson loop expectation values $\left\langle
W\right\rangle $ (red circles, dotted line, multiplied by $450$) along the
time direction (left panel) and along the spacial diagonals (right panel)
from the\ center of the simulation box for $g^{2}=1$. (The dashed horizontal
lines are fits to constant plateaus for $t,r\leq 1$. The two vertical lines
indicate the boundaries of the core and sampling volumes.)}
\label{sTg1fig}
\end{figure}

\begin{figure}[tbp]
\begin{center}
\includegraphics[height = 7cm]{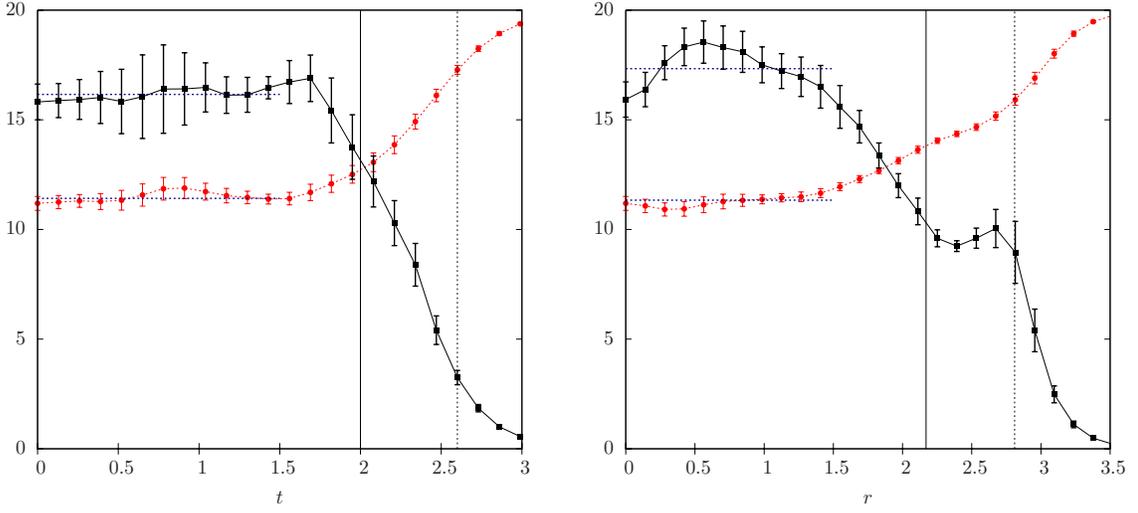}
\end{center}
\caption{Same as in Fig. \protect\ref{sTg1fig}, but for $g^{2}=25$ (and with 
$\left\langle W\right\rangle $ multiplied by 20).}
\label{sDg5fig}
\end{figure}

The action density $s\left( x\right) $, as defined in Eq. (\ref{sym}), can
be obtained analytically by inserting the dimeron superposition (\ref{cf})
into the field strength (\ref{fym}). The ensemble average $\left\langle
s\right\rangle $ is then evaluated as outlined in Sec. \ref{mcsec}, with
ensemble configurations taken from every fifth sweep after equilibration. In
order to visualize the local homogeneity properties of the action density in
different sectors of the simulation volume, we plot $\left\langle
s\right\rangle $ in Figs. \ref{sTg1fig} and \ref{sDg5fig} as a function of
the distance from the center of the box along two different directions.\ The
latter were chosen to provide insight into the anisotropy of the boundary
effects, caused in particular by our asymmetric grid with its elongated
temporal direction. (Instead of calculating the values of $s\left( x\right) $
at neighboring, equidistant points throughout these directions, we surround
them with adjacent hypercubic boxes of side length $0.2$ inside which we
sample $s\left( x\right) $ at 10 randomly chosen points.)

We start our discussion of the $\left\langle s\right\rangle $ plots with
those generated at our smallest coupling value, $g^{2}=1$. In the left panel
of Fig. \ref{sTg1fig} we show $\left\langle s\right\rangle $ (black squares,
full line)\ along the time direction, in the right panel along the average
over all diagonals of the spacial box at the mid point of the time axis.
Deviations from a translationally invariant, i.e. constant $\left\langle
s\right\rangle $ near the center of the box are somewhat smaller along the
time direction than along the diagonal. This is expected since the former
keeps a larger distance from the spacial\ boundaries. Nonetheless, $%
\left\langle s\right\rangle $ fluctuates considerably even in the temporal
direction and even close to the box center. A reasonable fit to a constant
plateau remains possible inside most error bars up to distances of order $%
1.0 $ both along the time and diagonal directions, however, as also shown in
Fig. \ref{sTg1fig}. Beyond distances of about ten lattice units the missing
field contributions from outside of the box volume (cf. Sec. \ref{box})
begin to reduce the action density substantially.

As a consequence of these results, we will restrict the volume in which we
evaluate amplitudes to subvolumes of the core box\ in which the maximal
distances from the center are of order one (if not noted otherwise). To
demonstrate the $g^{2}$ dependence of the boundary effects, we further show
the two analogous $\left\langle s\right\rangle $ profiles for $g^{2}=25$ in
Fig. \ref{sDg5fig}. Clearly, the boundary effects are strongly reduced and
the regions close to the center now show manifest plateaus which extend up
to distances $\sim 1.5$ from the center. With further increasing $g^{2}$ the
boundary artefacts become even more restricted to the surface region until
the growing dimeron dissociation (i.e. the growing intermeron distance $%
2\left\vert a\right\vert $) creates a different type of sensitivity to the
boundary, as discussed in Secs. \ref{box} and \ref{dimenssec}.

Since in general both character and strength of the boundary artefacts are
amplitude dependent, we have also monitored the vacuum expectation value of
quadratic \textquotedblleft probe\textquotedblright\ Wilson loops $W$ as a
function of their distance from the box center. The results will guide us in
reliably calculating the expectation values of rectangular Wilson loops in
Sec. \ref{wlsec}. Since the largest among these loops play a crucial role in
understanding the confinement properties of dimeron ensembles, it becomes
especially important to understand the boundary's impact on them. We survey
it by evaluating the expectation values $\left\langle W\right\rangle $ of
quadratic Wilson loops (for details see Sec. \ref{wlsec}) of side length $%
0.5 $ centered along the same two lines as the action density $\left\langle
s\right\rangle $ above. The loop orientation turns out to have no
significant impact on the value of $\left\langle W\right\rangle $ and will
thus be averaged over. In Fig. \ref{sTg1fig} we have included plots of $%
450\times \left\langle W\right\rangle $ along the time direction in the left
panel and along the average over the spacial diagonals in the right panel
for $g^{2}=1$, and in Fig. \ref{sDg5fig} we display $20\times \left\langle
W\right\rangle $ for $g^{2}=25$. In all four cases, with growing $t,r$ the
distance dependence of $\left\langle W\right\rangle $ becomes increasingly
proportional to that of the average action density $\left\langle
s\right\rangle $. For quadratic Wilson loops of minimal size this is
expected since such \textquotedblleft plaquettes\textquotedblright\ form the
essential part of a discrete approximation to the Yang-Mills action density.
In our case, the observed proportionality further indicates that the
configurations are sufficiently smooth over our rather large test loops,
probably because their side length is comparable to the regulated meron size 
$\rho \simeq 0.55$.

To summarize, while discretization errors in our dimeron simulations can be
reliably controlled, we have observed significant violations of
translational invariance in both action density and Wilson loops. Although
our efforts to reduce boundary artefacts prove effective close to the center
of the core box, substantial deviations from homogeneity set in at distances
of order one, in particular for our smallest coupling value $g=1$ where they
are most pronounced. These boundary effects originate from a combination of
the strong intermediate-range interactions between dimerons and the
collective effects emanating from the boundary (cf. Sec. \ref{box}). The
lessons learned from the above analysis will guide us in keeping boundary
artefacts of our results tolerably small, mainly by relying on specifically
reduced volumes in which to evaluate amplitudes.

\subsection{Ensemble statistics}

\label{statsec}

The effective dimeron field theory introduced in Sec. \ref{sdeft} has three
adjustable parameters: the gauge coupling $g^{2}$, the meron-center size $%
\rho $, and the approximately equal numbers $N_{D}\simeq N_{\overline{D}}$
of dimerons and antidimerons. We have generated all our ensembles with $%
N=N_{D}+N_{\overline{D}}=487$ pseudoparticles, divided into $N_{D}=243\simeq
N/2$ dimerons and $N_{\overline{D}}=244\simeq N/2$ antidimerons, in the
multi-layered multi-grid volume specified in Sec. \ref{box}. The meron
centers are smeared as described in Sec. \ref{reg}, with a common regulator
value $\rho ^{2}=0.3$. We have generated two independent ensembles for each
of the coupling values $g^{2}\in \left\{ 1,25,100,1000,\infty \right\} $
(with decreasing number of members) and a third one for $g^{2}=1$. The total
number of sweeps per coupling in equilibrium as well as the maximal
stepsizes are collected in Tab. \ref{enstab}

%TCIMACRO{\TeXButton{B}{\begin{table}[htbp] \centering}}%
%BeginExpansion
\begin{table}[htbp] \centering%
%EndExpansion
\begin{tabular}{|c|c|c|c|}
\hline
$g^{2}$ & $\#$ of sweeps & $\left\vert da\right\vert _{\max }$ & $\left\vert
dc\right\vert _{\max }$ \\ \hline\hline
\multicolumn{1}{|l|}{$1$ \ \ \ \ \ \ } & \multicolumn{1}{|c|}{$1843$} & $%
0.04 $ & $0.11$ \\ \hline
\multicolumn{1}{|l|}{$25$} & \multicolumn{1}{|c|}{$1550$} & $0.20$ & $0.40$
\\ \hline
\multicolumn{1}{|l|}{$100$} & \multicolumn{1}{|c|}{$1470$} & $0.36$ & $0.44$
\\ \hline
\multicolumn{1}{|l|}{$1000$} & \multicolumn{1}{|c|}{$554$} & $0.47$ & $0.49$
\\ \hline
\multicolumn{1}{|l|}{$\infty $} & \multicolumn{1}{|c|}{$574$} & $0.10$ & $%
1.00$ \\ \hline
\end{tabular}%
\caption{Dimeron ensemble characteristics. (For definitions see Sec. \ref{mcsec}.)
\label{enstab}} 
%TCIMACRO{\TeXButton{E}{\end{table}}}%
%BeginExpansion
\end{table}%
%EndExpansion

In order to estimate autocorrelation effects in the ensembles with $g^{2}=1$%
, where they should be maximal, we have calculated autocorrelation functions
for several quantities of interest, including the average distance between
nearest-neighbor meron centers of equal and opposite topological charges.
For those of opposite topological charge the autocorrelations decay fastest,
after about 300 -- 400 equilibrium sweeps, while this takes up to a few
hundered sweps longer for equally charged neighbors. The autocorrelation
functions for the average distance between nearest-neighbor dimerons, on the
other hand, show no obvious preference for either equally or oppositely
charged neighbors. Both distances approximately decorrelate after 500
sweeps. The same holds for the average distance between the dimerons' meron
partners.

Based on the above observations, we have adopted the following strategy to
reduce autocorrelations. The ensembles contain the configurations generated
by every fifth Metropolis sweep. The total number of equilibrium sweeps (per
coupling) is divided into ten bins. (For $g^{2}=1$ a bin thus contains about 
$180$ sweeps and about $36$ ensemble configurations.) The quantities of
interest are then calculated on every ensemble configuration, and their mean
values are obtained for each bin. Finally, the ensemble average is computed
as the mean value of the bin averages, and its error is estimated as the
standard deviation among the bin averages (if not noted otherwise). (On an
Intel Core 2 Quad Q6700 processor at 2.67 Ghz a thermal Markov step takes
about $t_{m}\sim 7$ seconds and a thermal sweep $2Nt_{m}\sim \allowbreak
1.\,\allowbreak 89$ hours.)

\section{Results and discussion}

\label{ressec}

In the following subsections we analyze the physics content of dimeron
ensembles which were numerically generated at the five squared coupling
values $g^{2}=\left\{ 1,25,100,1000,\infty \right\} $ according to the
procedure outlined in Sec. \ref{simul}.

\subsection{Dimeron dissociation as a function of the gauge coupling}

\label{dissocsec}

One of the essential features of the CDG mechanism is that with increasing
coupling instantons are supposed to gradually dissociate into dimerons and
to finally break up into their two meron partners. This process is suggested
to be driven by a competition between the attraction among the regularized
meron centers and their position entropy which increases with $g^{2}$ (since
the latter plays the role of temperature in the classical statistical analog
ensemble). The competition is effective because both energy and entropy
depend logarithmically on the distance $2\left\vert a\right\vert $ between
the meron centers. In fact, this distance will play a key role in
characterizing both the structure of the ensembles' dimeron constituents and
the interactions among them. The latter depend sensitively on the dimerons'
color dipole moment which is of $O\left( a^{2}\right) $.

Before the dimerons break up completely, they should already have
effectively released their two meron centers which then become the
dynamically active degrees of freedom. In fact, this is expected to signal
the onset of a phase transition in a finite system like ours where a
complete $\left\vert a\right\vert \rightarrow \infty $ dissociation is
prevented by the boundary. With their slowly decaying and thus strongly
overlapping long-distance tails ($\sim 1/x$) the essentially independent
merons may then sufficiently disorder the vacuum to generate linear quark
confinement. In the above sense, our dimeron configurations thus approach
confining meron ensembles of the type studied in Refs. \cite{len08,len04}.
One should keep in mind, however, that our dimeron superposition ansatz (\ref%
{cf}) is not rich enough to describe ensembles of completely independent
merons since it does not allow to untie the rigid color locking between the
meron partners.

\begin{figure}[tbp]
\begin{center}
\includegraphics[height = 7cm]{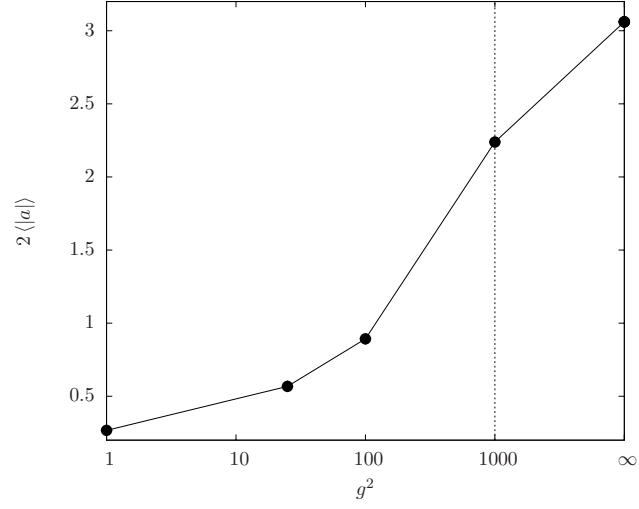}
\end{center}
\caption{The average separation $2\left\langle \left\vert a\right\vert
\right\rangle $ between the (anti-) meron centers of the (anti-) dimerons as
a function of the squared gauge coupling. (The error bars are smaller than
the plot symbols and the dotted vertical line indicates the scale change at $%
g^{2}=1000$.)}
\label{dissocfig}
\end{figure}

Guided by the above considerations we are thus led to study the coupling
dependence of the average dimeron dissociation $2\left\langle \left\vert
a\right\vert \right\rangle $ in our ensembles. These average inter-meron
distances, which contribute to the typical dimeron size scale\ $\sim 2\left(
\rho +\left\langle \left\vert a\right\vert \right\rangle \right) $, are
plotted as a function of the square coupling $g^{2}$ in Fig. \ref{dissocfig}
and listed in Table \ref{dissoctab}. As expected, the average dimeron
dissociation $2\left\langle \left\vert a\right\vert \right\rangle $
initially increases rather strongly with $g^{2}$. For our two largest square
couplings $g^{2}=1000$ and $\infty $, on the other hand, its value
approaches saturation since it becomes comparable to the linear extent of
the ensemble box. Hence $2\left\langle \left\vert a\right\vert \right\rangle 
$ will be underestimated in this coupling region. Moreover, large dimerons\
close to the boundary are then increasingly forced to align themselves with
the boundaries, and the largest dimerons accumulate along the box diagonals.
Below we will nevertheless find indications for the meron centers (which are
color zero-poles) to gradually replace the dimerons (i.e. color dipoles) as
the dynamically most relevant degrees of freedom when the coupling
increases, as alluded to above.

%TCIMACRO{\TeXButton{B}{\begin{table}[htbp] \centering}}%
%BeginExpansion
\begin{table}[htbp] \centering%
%EndExpansion
\begin{tabular}{|c|c|}
\hline
$g^{2}$ & $2\left\langle \left\vert a\right\vert \right\rangle $ \\ 
\hline\hline
\multicolumn{1}{|l|}{$1$ \ \ \ \ \ } & $0.267\pm 0.003$ \\ \hline
\multicolumn{1}{|l|}{$25$} & $0.567\pm 0.001$ \\ \hline
\multicolumn{1}{|l|}{$100$} & $0.893\pm 0.003$ \\ \hline
\multicolumn{1}{|l|}{$1000$} & $2.239\pm 0.014$ \\ \hline
\multicolumn{1}{|l|}{$\infty $} & $3.061\pm 0.016$ \\ \hline
\end{tabular}%
\caption{The average inter-meron distance $2 \langle |a| \rangle$ for five values 
of the square coupling.
\label{dissoctab}} 
%TCIMACRO{\TeXButton{E}{\end{table}}}%
%BeginExpansion
\end{table}%
%EndExpansion

\subsection{Spacetime structure of the dimeron configurations}

\label{dimenssec}

We begin to explore the multi-dimeron physics of our ensembles by analyzing
the spacetime structure of a typical member configuration. To this end we
survey the action density of this field configuration and its amount of
local (anti-) selfduality on different hyperplanes of the simulation volume.
More specifically, we display these quantities as two-dimensional density
plots in two section planes through the core volume which are chosen
parallel to the $x_{4}-x_{1}$ plane. One of them is separated from the $%
x_{4}-x_{1}$ plane by the distances $\left( x_{2},x_{3}\right) =\left(
0.4,0.4\right) $ in the remaining directions, i.e. it lies rather close to
the boundary in these transverse directions. The other one is positioned
close to the center in the $x_{3}$ direction, at $\left( x_{2},x_{3}\right)
=\left( 0.4,1.2\right) $. (The origin of our coordinate system coincides
with a corner of the core box. The spacial axes $x_{1,2,3}$ range from $0$
to $2.5$ and the $x_{4}$ axis from $0$ to $4$. Hence the box center is
located at $\left( x_{1},x_{2},x_{3},x_{4}\right) =\left(
1.25,1.25,1.25,2\right) $, cf.. Sec. \ref{box}.)

Since the detection and control of boundary effects is a recurrent issue
when dealing with dimeron ensembles, we draw all plots for a $g^{2}=1$
configuration where the impact of the boundary is maximal (at least up to $%
g^{2}\simeq 1000$) and thus best analyzable. Such weak-coupling
configurations with their small entropy are of additional interest because
they are most likely to approximate semi-classical fields. In fact, at not
too high densities (and not too large meron size regulators) such fields
would be dominated by rather isolated and strongly contracted singular-gauge
dimerons. These dilute superpositions of almost instanton-like solutions
(cf. Sec. \ref{dissocsec}) indeed approximate semiclassical systems quite
similar to those studied in instanton vacuum models \cite{sch98}. The
following plots are designed to check how far our $g^{2}=1$ configuration
resembles such semi-classical fields.

Although our numerically generated dimeron field configurations contain
redundant, i.e. gauge-dependent information without impact on observables,
it is of technical interest to understand their spacetime structure since
they lay the foundation for all our ensuing work. We have therefore examined
typical dimeron ensemble configurations on the above set of hypersurfaces
and found the fields to be fairly smooth. More importantly, all prominent
spacial features of the gauge fields components were found to be closely
mirrored in the gauge-invariant densities to be discussed below. In
particular, we have found no evidence for the build-up of gauge-dependent
peaks (potentially approximating gauge singularities \cite{def01}) which
could adversely affect the simulation behavior even though they are
invisible in gauge-invariant quantities. Hence we can refrain from plotting
selected components of the gauge field itself.

\begin{figure}[tbp]
\begin{center}
\includegraphics[height = 10cm]{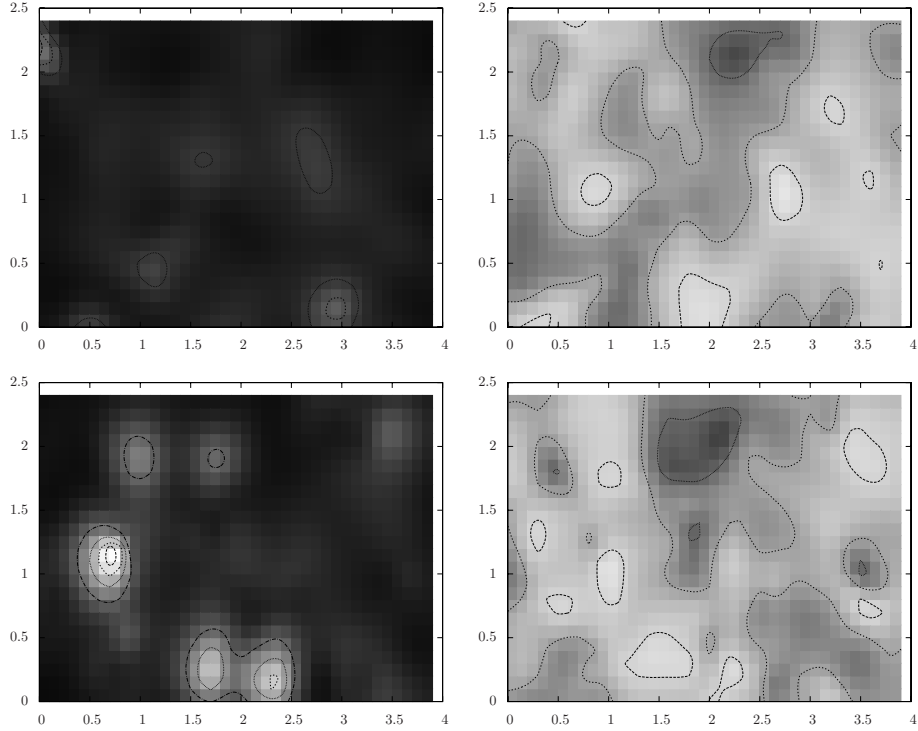}
\end{center}
\caption{The action (left panels) and selfduality (right panels) densities
of a typical $g^{2}=1$ ensemble configuration in two cross sections parallel
to the $x_{1}-x_{4}$ plane. Lighter shades of gray indicate larger values.\
The ordinates denote the $x_{1}$ and the abcissas the $x_{4}$ direction. The
upper row shows the densities in the hyperplane at $\left(
x_{2},x_{3}\right) =\left( 0.4,0.4\right) $ and the lower row those in the
plane at $\left( x_{2},x_{3}\right) =\left( 0.4,1.2\right) $. }
\label{acdenssdfig}
\end{figure}

Instead, we turn to the gauge-invariant action density $s\left( x\right) $
of our example configuration which is plotted in the two mentioned planes
through the core volume in the two left panels of Fig. \ref{acdenssdfig}.
The graphs indicate that the action density is indeed rather smooth, with
the exception of a few dilute peaks. All these peaks show a slightly
nonspherical shape and an extension of about $0.5$. Now we recall from Table %
\ref{dissoctab} that for $g^{2}=1$ the average distance between the (anti-)
meron centers of the (anti-) dimerons is $2\left\langle \left\vert
a\right\vert \right\rangle \sim 0.27$. This is about half of the regularized
meron size $\rho \simeq 0.55$ and suggests to identify the peaks with the
two strongly overlapping meron centers of the regularized dimerons. The
average number of peaks is indeed consistent with the pseudoparticle density
of the configuration. Moreover, the somewhat elongated action density of the
peaks finds a natural explanation in the small but finite separation between
the meron centers. Finally, in the more central $\left( x_{2},x_{3}\right)
=\left( 0.4,1.2\right) $ plane the peak density is larger than in the $%
\left( x_{2},x_{3}\right) =\left( 0.4,0.4\right) $ plane which lies closer
to the boundary. This may be a reflection of the strongly reduced average
action density near the boundary (cf. Sec. \ref{eran} and Fig. \ref{sTg1fig}%
). Statistical fluctuations are too large to substantiate this conjecture,
however, as indicated by the fact that no such dilution is recognizable in
Fig. \ref{acdenssdfig} close to the boundaries in the $x_{1},x_{4}$
directions.

Another instructive property of the dimeron field configurations is their
amount of local selfduality. This feature characterizes the interplay
between the Yang-Mills dynamics and the topological charge density $q\left(
x\right) $ (as defined in Eq. (\ref{q})) and can be monitored by evaluating
the expression 
\begin{equation}
R\left( x\right) =\frac{4}{\pi }\arctan \left( \sqrt{\frac{tr\left\{ F_{\mu
\nu }F_{\mu \nu }-F_{\mu \nu }\widetilde{F}_{\mu \nu }\right\} }{tr\left\{
F_{\mu \nu }F_{\mu \nu }+F_{\mu \nu }\widetilde{F}_{\mu \nu }\right\} }}%
\right) -1
\end{equation}%
which varies from $-1$ at positions where the field is selfdual to $+1$
where it is anti-selfdual. Since for $g^{2}=1$ the strongly overlapping,
regularized meron centers render the (anti-) dimerons almost (anti-)
instanton-like, one expects that the peaks are approximately (anti-)
selfdual while the surrounding regions, dominated by overlapping tails, are
neither. The plots of $R\left( x\right) $ in the right panels of Fig. \ref%
{acdenssdfig} confirm this expectation and thus allow to associate the
action density peaks with either dimerons or anti-dimerons. Due to the
one-to-one correspondence between the peaks in $s\left( x\right) $ and $%
R\left( x\right) $ the above comment on boundary effects applies here as
well.

\subsection{Topological charge distribution}

\label{topdissec}

\begin{figure}[tbp]
\begin{center}
\includegraphics[height = 6cm]{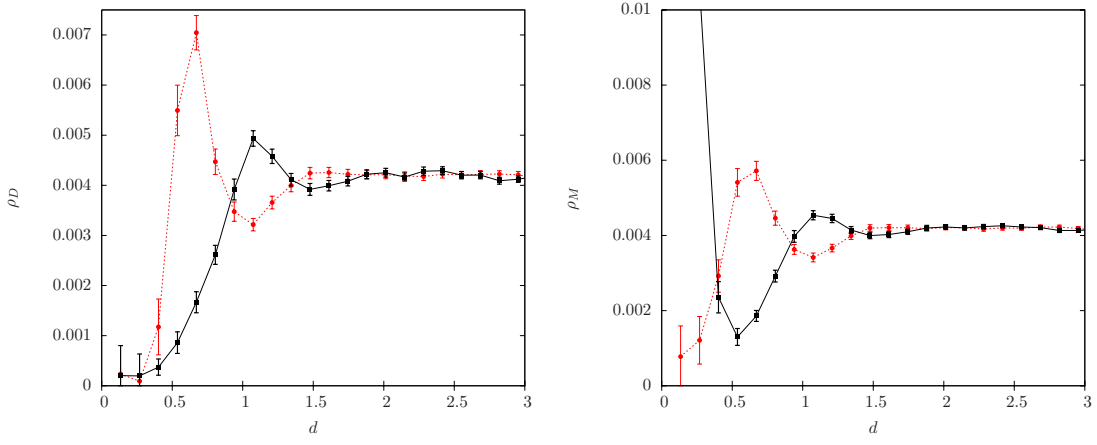}
\end{center}
\caption{The average radial density of topologically equally (black squares,
full line) and oppositely (red bullets, dotted line) charged dimerons $%
\protect\rho _{D}$ (left panel) and merons $\protect\rho _{M}$ (right panel)
as a function of the distance from a fixed (anti-) (di)meron at $g^{2}=1$.}
\label{dimerdens1fig}
\end{figure}

\begin{figure}[tbp]
\begin{center}
\includegraphics[height = 6cm]{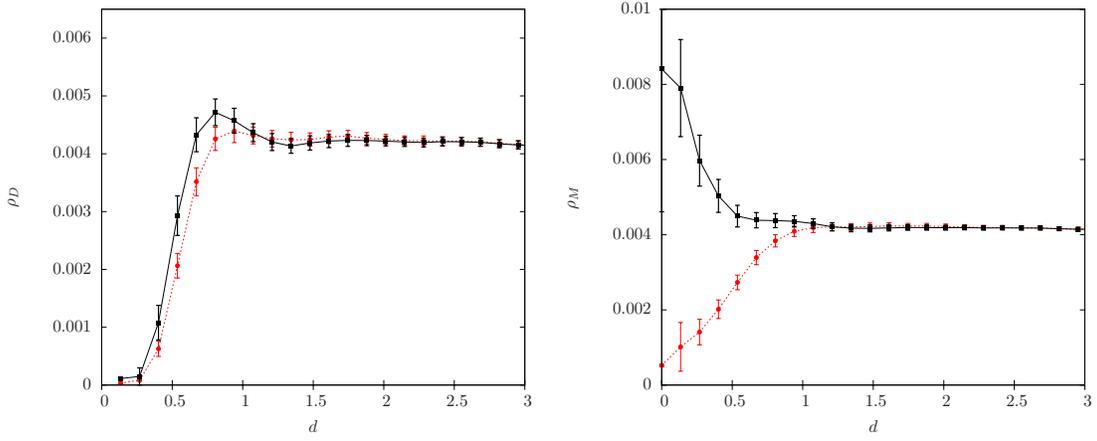}
\end{center}
\caption{The same as in Fig. \protect\ref{dimerdens1fig}, but for $g^{2}=25$%
. }
\label{dimerdens5fig}
\end{figure}

\begin{figure}[tbp]
\begin{center}
\includegraphics[height = 6cm]{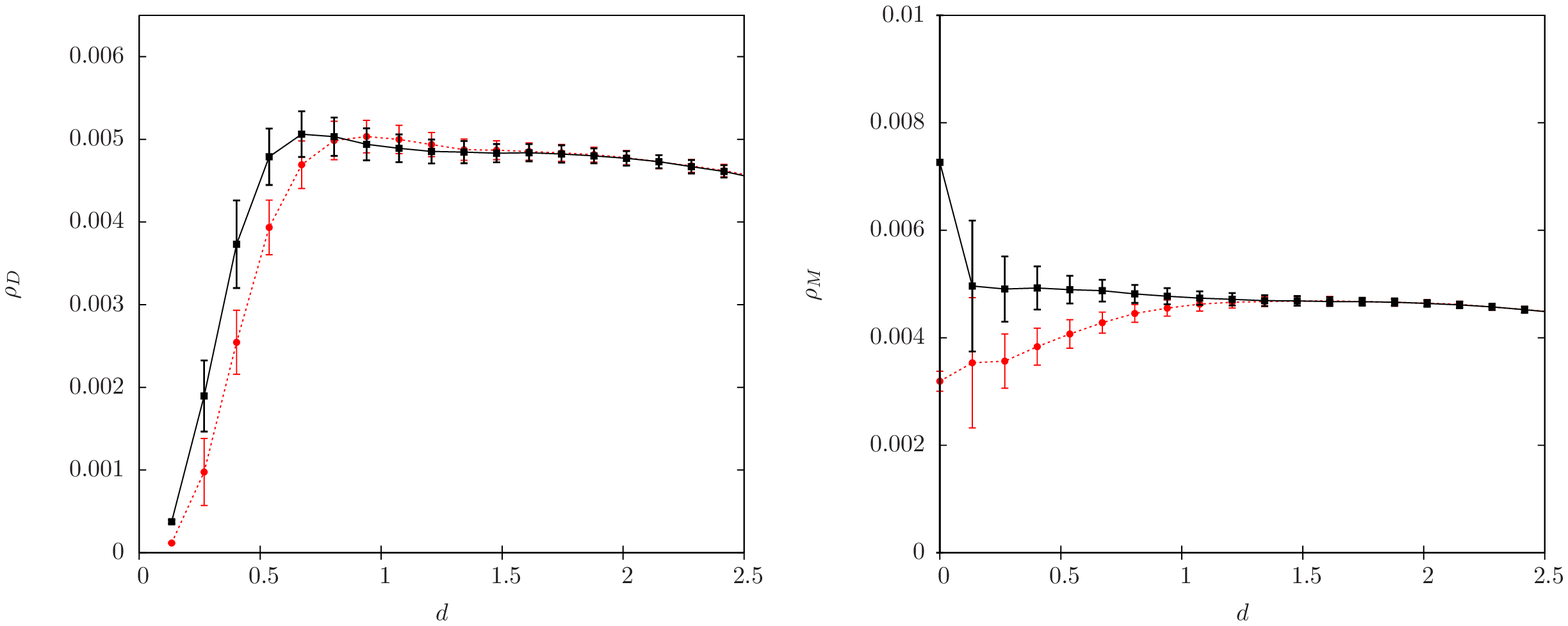}
\end{center}
\caption{The same as in Fig. \protect\ref{dimerdens1fig}, but for $g^{2}=100$%
.}
\label{dimerdens10fig}
\end{figure}

We now proceed to a more quantitative analysis of the dimeron ensemble
structure and its dependence on the square coupling $g^{2}$. In the present
section we look for patterns in the topological charge density which
indicate short- and medium-range order. More specifically, we pick a
pseudoparticle in a given ensemble configuration and measure the average
density of the surrounding pseudoparticles with either equal or opposite
topological charge, as a function of their distance from the selected one 
\footnote{%
In practice, these densities were calculated by dividing the box volume into
60 concentric, spherical shells around the reference pseudoparticle,
counting the number of either equally or oppositely charged pseudoparticle
centers contained in it, and dividing by the shell volume $V_{\text{shell}}$%
. The latter is calculated analytically as long as the shells lie fully
inside the ensemble volume (cf. Sec. \ref{box}), and otherwise as $V_{\text{%
shell}}=VN_{\text{shell}}/N_{V}$ where $N_{V}$ points are randomly
distributed over the ensemble box and $N_{\text{shell}}$ is the number of
them which fall into the shell.}. We then repeat this procedure for all
pseudoparticles in the configuration and average over the obtained\ density
profiles, again for equal and opposite topological charge separately. After
finally averaging over all configurations of the $g^{2}=1$ ensemble we end
up with the two distance profiles $\rho _{D}$ plotted in the left panel of
Fig. \ref{dimerdens1fig}. The analogous procedure, but with the
pseudoparticles replaced by individual meron centers, yields the profiles $%
\rho _{M}$ shown in the right panel. The integral of these densities over
the full ensemble is normalized to one. (The smallest and largest distances
are excluded in both figures since they correspond to tiny shell volumes
(inside the ensemble volume) in which the densities cannot be reliably
estimated.) Figures \ref{dimerdens5fig} and \ref{dimerdens10fig} contain the
same profiles as Fig. \ref{dimerdens1fig}, but for $g^{2}=25$ and $g^{2}=100$%
.

The averaged radial density profiles reveal an intriguing amount of
structure in the pseudoparticle distribution and in its meron substructure.
First of all, the left panels of Figs. \ref{dimerdens1fig} -- \ref%
{dimerdens10fig} show a depletion of both dimeron and anti-dimeron densities
in the overlap region with the reference (anti-)dimeron. Hence they provide
direct evidence for a strong short-distance repulsion between dimerons of
any topological charge. This repulsion is sensitive to the relative color
arrangement between neighboring pseudoparticles (cf. Sec. \ref{colcorsec})
and may\ at least partly be caused by our limited field configuration space.
A similar repulsive core (with a logarithmic distance dependence) shows up
in superpositions of instantons and anti-instantons in singular gauge \cite%
{dia84}. On a practical level, it helps to avoid clustering among the
pseudoparticles and promotes smoother and more semiclassical ensemble
configurations. Evidence for the latter was already encountered in the
density plots of Sec. \ref{dimenssec}.

At larger distances, a remarkable medium-range order emerges among the
pseudoparticles for $g^{2}=1$ (left panel of Fig. \ref{dimerdens1fig}). In
fact, about one meron size $\rho \sim 0.55$ (or the slightly larger average
dimeron size $\rho +\left\langle \left\vert a\right\vert \right\rangle $)
from the fixed reference particle one finds an enhanced (depleted)\ density
of pseudoparticles with opposite (equal) topological charge. At about $2\rho 
$ this structure is inverted and attenuated, i.e. the density of
pseudoparticles with equal (opposite) charge is weakly enhanced (depleted).
A further, even weaker inversion of the densities is discernible at
distances $\sim 3\rho $, while from $d\gtrsim 3.5\rho $ both dimeron and
antidimeron densities remain within errors equal to those of a random
distribution. The almost periodic density oscillations over three
consecutive layers indicate a pronounced mid-range order among the dimerons.
In fact, the emerging shell structure resembles Debye-type screening clouds
and indicates the existence of attractive short-distance correlations
between dimerons and anti-dimerons \footnote{%
This\ screening behavior should not be confused with the light-quark and
anomaly-induced topological charge screening in the QCD vacuum \cite{chu00},
with its strong impact on ${\protect \eta}^{\prime}$ meson \cite{dow92} and
pseudoscalar glueball properties \cite{for205,shu95}.}.

The above screening behavior should be enhanced at $g^{2}=1$ where the
entropy is lowest and the field configurations therefore most strongly
ordered. This can indeed be seen in our results. While for $g^{2}=1$ the
third shells are clearly visible in Fig. \ref{dimerdens1fig} (although less
pronounced than the first two), they essentially disappear for $g^{2}\geq 25$
(cf. Figs. \ref{dimerdens5fig} and \ref{dimerdens10fig}). Moreover, the
dimeron densities in the left panels of Figs. \ref{dimerdens5fig} and \ref%
{dimerdens10fig} show a weaker first shell at somewhat larger distances
(reflecting the growing average size of the dimerons), which now slightly
favors pseudoparticles of equal topological charge. A hint of a second shell
with inverted topological charge remains recognizable as well. Hence at
stronger coupling and over typical nearest-neighbor distances the (anti-)
dimerons show a tendency to surround themselves with (anti-) dimerons. This
behavior could be another indication for the with growing dimeron
dissociation increasing role of the meron centers as the dynamically
relevant degrees of freedom.

In order to test the latter interpretation, we show in the right panels of
Figs. \ref{dimerdens1fig} -- \ref{dimerdens10fig} the analogous density
profiles $\rho _{M}$ for individual meron centers (without regard for the
dimerons which they are part of). At the smallest distances $d\ll \rho $,
i.e. in the immediate overlap region with the fixed reference meron, the
nearest-neighbor meron is most likely its partner from the same dimeron, and
hence has the same topological charge. This explains the enhanced
(suppressed) density of equally (oppositely) charged merons for $d\ll \rho $%
. The enhancement is maximal for $g^{2}=1$ (cf. Fig. \ref{dimerdens1fig})
where the mean inter-meron distance $2\left\langle \left\vert a\right\vert
\right\rangle $ is minimal, and decreases due the growing $\left\langle
\left\vert a\right\vert \right\rangle $ with increasing $g^{2}$. Outside of
the immediate overlap region, on the other hand, i.e. for distances $d>\rho $%
, the meron density profiles in the right panels of Figs. \ref{dimerdens1fig}
-- \ref{dimerdens10fig} follow those of the corresponding dimeron densities
(left panels) rather closely. This is expected because for $d>\rho $ the
encountered merons belong more likely to different dimerons. Since for $%
g^{2}=1$ with $2\left\langle \left\vert a\right\vert \right\rangle \simeq
0.27\sim \rho /2$ the meron partners of the dimerons overlap almost
completely, furthermore, their densities outside the immediate neighborhood
of the reference meron should match those of the dimerons most closely, as
is indeed the case.

\begin{figure}[tbp]
\begin{center}
\includegraphics[height = 7cm]{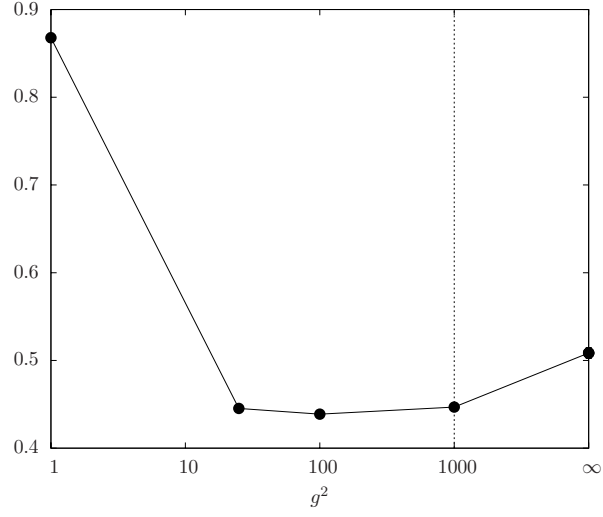}
\end{center}
\caption{The average probability $\left\langle f_{\text{D}\overline{\text{D}}%
}\right\rangle $ for the nearest neighbor of a dimeron to have opposite
topological charge. (The error bars are smaller than the plot symbols and
the dotted vertical line indicates the scale change at $g^{2}=1000$.)}
\label{proboppchargefig}
\end{figure}

We have already pointed out that the strongest intermediate-range order
among the dimerons exists in the $g^{2}=1$ ensemble with its particularly
low entropy. For a more systematic analysis of the changes in this behavior
with increasing coupling we now proceed to the investigation of global
ensemble properties. (Their smaller statistical error simplifies the study
of the $g^{2}$ dependence.) We first consider the fraction $f_{\text{D}%
\overline{\text{D}}}$ of dimerons in a given configuration whose nearest
neighbor has opposite topological charge \footnote{%
In order to reduce the impact of boundary effects we include only those
(anti-)dimerons whose centers lie inside a (hyper)sphere of radius one
around the mid point of the box. The inter-meron separation $2\left\langle
\left\vert a\right\vert \right\rangle $ of the dimerons then restricts the
maximal distance of the meron center from the box center to $1+\left\langle
\left\vert a\right\vert \right\rangle $, i.e. on average meron centers lie
maximally a distance $\left\langle \left\vert a\right\vert \right\rangle $
outside the ball. For $g^{2}=1,$ where the boundary effects are strongest,
the smallest $\left\langle \left\vert a\right\vert \right\rangle \sim 0.27$
then reduces their impact by ensuring that the meron centers remain closest
to the sphere. For the maximal couplings, on the other hand, one finds $%
\left\langle \left\vert a\right\vert \right\rangle \sim 3/2$.}. The
ensemble-averaged probability $\left\langle f_{\text{D}\overline{\text{D}}%
}\right\rangle $ is plotted in Fig. \ref{proboppchargefig} for our five $%
g^{2}$ values between one and infinity. (The vertical line indicates a break
in the scale of the abscissa which allows to include the strong-coupling
limit $g^{2}=\infty $.) For $g^{2}=1$ one reads off $\left\langle f_{\text{D}%
\overline{\text{D}}}\right\rangle \simeq 87$\%. This large fraction confirms
the strong preference of the dimerons to surround themselves with screening
clouds consisting mostly of their anti-pseudoparticles. Already at $g^{2}=25$
the value of $\left\langle f_{\text{D}\overline{\text{D}}}\right\rangle $
has diminished by half, however, to about 43\%. This is a clear indication
for the with increasing coupling growing disorder in the field
configurations. It manifests itself not the least in the stronger
dissociation of the dimerons ($2\left\langle \left\vert a\right\vert
\right\rangle \simeq 0.57\simeq \rho $ at $g^{2}=25,$ cf. Table \ref%
{dissoctab}) which gradually replaces the interactions between dimeron
centers by interactions between the increasingly independent meron centers.%
\textbf{\ }The fact that $\left\langle f_{\text{D}\overline{\text{D}}%
}\right\rangle $ remains within errors around 45\% for $g^{2}=100$ and $1000$
signals the slight preference for equally charged pseudoparticle neighbors
at stronger couplings, as already encountered in Figs. \ref{dimerdens5fig}
and \ref{dimerdens10fig}. At these couplings the dimerons are so far
dissociated, however, that correlations between their centers do probably no
longer characterize the main interactions which they experience. In any
case, the non-interacting value of 50\% for $\left\langle f_{\text{D}%
\overline{\text{D}}}\right\rangle $ is attained only in the strong-coupling
limit, i.e. in the random ensemble with $g^{2}=\infty $ (notwithstanding
additional boundary effects which come into play for maximally dissociated
dimerons, cf. Sec. \ref{eran}).

\begin{figure}[tbp]
\begin{center}
\includegraphics[height = 7cm]{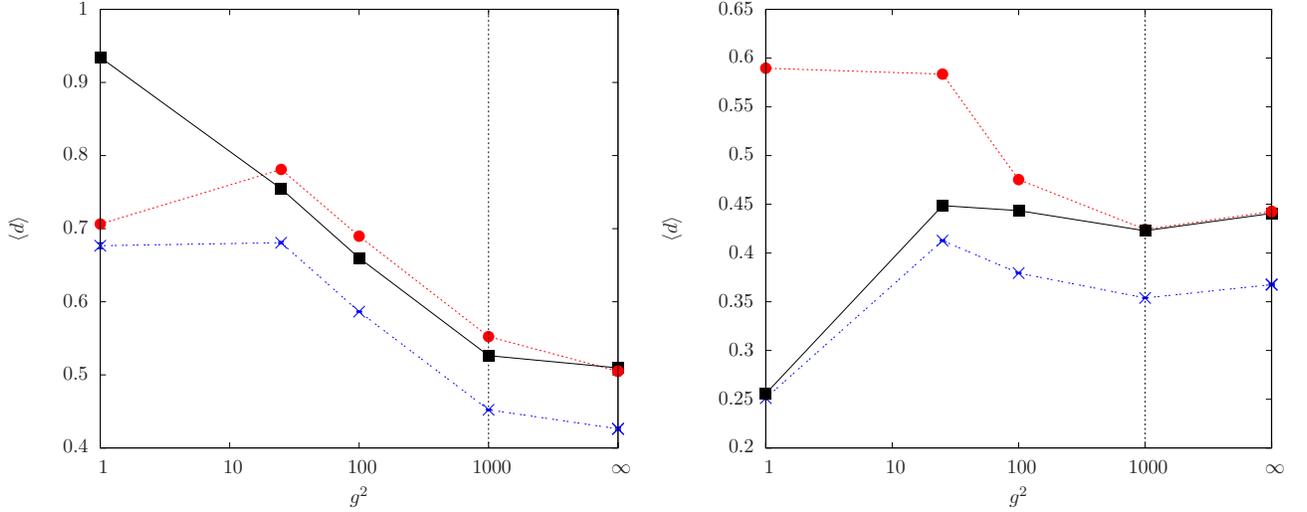}
\end{center}
\caption{The left panel shows the average distance $\left\langle
d\right\rangle $ of a reference dimeron center from its nearest neighbor
with equal (black squares, full line) and opposite (red bullets, dotted
line) topological charge $Q$. The nearest-neighbor distance $\left\langle
d\right\rangle $ independent of $Q$ is also included (blue crosses, dashed
line). The right panel shows the same curves for the average distances
between meron instead of dimeron centers. (The error bars are smaller than
the plot symbols and the dotted vertical line indicates the scale change at $%
g^{2}=1000$.)}
\label{ndistfig}
\end{figure}

Further insight into the topological charge distribution of dimeron
ensembles and its coupling dependence can be obtained from the average
distances $\left\langle d\right\rangle $ between the pseudoparticle centers
and their nearest neighbors with either equal or opposite topological
charge. The nearest-dimeron distances are drawn in the left panel of Fig. %
\ref{ndistfig} as a function of the square coupling $g^{2}$. (They are
evaluated in the same spherical shells as in Fig. \ref{proboppchargefig}.)
For $g^{2}=1$ one finds the average distance $\left\langle d\right\rangle
\simeq 0.71$ between oppositely charged neighbors to be about $25\%$ smaller
than that between equally charged ones ($\left\langle d\right\rangle \simeq
0.93$). This is a reflection of the screening clouds found above. After
increasing the square coupling to $g^{2}=25,$ on the other hand, the two
distances have become almost identical. The nearest pseudoparticles of equal
charge are now somewhat closer, as expected from the drop of $\left\langle
f_{\text{D}\overline{\text{D}}}\right\rangle $ below 50\%. (The average
distance to the next dimeron of any charge remains relatively constant,
though.) This seems to be another consequence of the larger inter-meron
separation $\left\langle 2\left\vert a\right\vert \right\rangle \simeq
0.6-3.1$ for $g^{2}=25-\infty $ and the increasing dynamical importance of
the meron centers. (For close-to-maximally dissociated dimerons the
mentioned boundary effects will distort the nearest-neighbor distances as
well, cf. Sec. \ref{eran}.)

Above we have found several pieces of evidence for the merons to gradually
turn into the dynamically relevant degrees of freedom when the coupling
increases. In order to explore this issue from yet another angle, we further
computed the average distances between nearest-neighbor meron centers, as
plotted in the right panel of Fig. \ref{ndistfig}. In the $g^{2}=1$ ensemble
with its highly contracted pseudoparticles, nearest-neighbor merons belong
most likely to the same dimeron and thus carry identical topological charge.
As a consequence, the average separation between equally charged neighbor
merons is almost the same as the distance $2\left\langle \left\vert
a\right\vert \right\rangle \simeq 0.27$ between partner merons. Oppositely
charged neighbor merons, on the other hand, are more than twice as far
separated, i.e. about as far as the average distance $\left\langle
d\right\rangle \simeq 0.59$ between nearest-neighbor dimerons of opposite
charge (cf. left panel of Fig. \ref{ndistfig}). Again, this picture changes
significantly with increasing coupling. As the dimerons dissociate farther,
the average distance $\left\langle d\right\rangle $ between equally charged
nearest-neighbor merons must increase as well. However, it does so more
slowly than the average separation $2\left\langle \left\vert a\right\vert
\right\rangle $ of the meron partners. This provides additional evidence for
the interaction between individual merons belonging to different dimerons to
increasingly determine the ensemble properties. At $g^{2}=25$ the average
distance between oppositely charged nearest-neighbor merons has not yet
decreased much, on the other hand. The main drop happens between $g^{2}=25$
and $100$. From about $g^{2}=1000$ the distances $\left\langle
d\right\rangle $ between equally and oppositely charged merons conincide and
slightly increase together on the approach to the strongly-coupled random
ensemble.

\subsection{Color correlations}

\label{colcorsec}

In the previous subsections we have found diverse spacial correlations in
dimeron ensemble configurations and studied their coupling dependence. It
remains to explore correlations between the color orientations of the
dimerons to which we turn in the present section. There are several reasons
for expecting these correlations to be significant. First, the
\textquotedblleft hedgehog\textquotedblright -type coupling between the
spacetime and color dependence of the individual dimerons (cf. Eq. (\ref%
{gmsing})) suggests that spacial and color correlations will be at least
partly linked in the ensemble configurations as well. In fact, we have
encountered a prototype of such correlations when studying the
color-orientation dependence of the classical interaction energy between two
isolated dimerons. In weakly coupled ensembles, furthermore, the almost
completely contracted dimerons will experience interactions and correlations
similar to those found in singular-gauge instanton ensembles (see below). In
the strong-coupling regime, on the other hand, where the dimerons are
dissociated and single merons with their long-range tails become dominant,
one may expect color correlations of similar strength and importance as in
the meron and regular-gauge instanton ensembles of Refs. \cite{len08,len04}.

In order to characterize the SU$\left( 2\right) $ color orientation between
two neighboring dimerons with color coordinates $c_{\beta }$ and $c_{\beta
}^{\prime }$ (cf. Eq. (\ref{quat})), we introduce the angle%
\begin{equation}
\alpha :=2\arcsin \left( \frac{1}{2}\sqrt{\left( c^{\prime }-c\right)
_{\beta }\left( c^{\prime }-c\right) _{\beta }}\right)  \label{alpha}
\end{equation}%
(with $\alpha \in \left[ 0,\pi \right] $) under which half of\ the geodesic
distance between the two corresponding points on the group manifold $S^{3}$
(of unit radius) appears from the center. We then find for every
pseudoparticle in a given ensemble configuration the nearest neighbors with
equal and opposite topological charges, separately calculate the angles (\ref%
{alpha}) between them and finally average over the configuration (excluding
double-counting of pairs) and over the $g^{2}=1$ ensemble.

\begin{figure}[tbp]
\begin{center}
\includegraphics[height = 8cm]{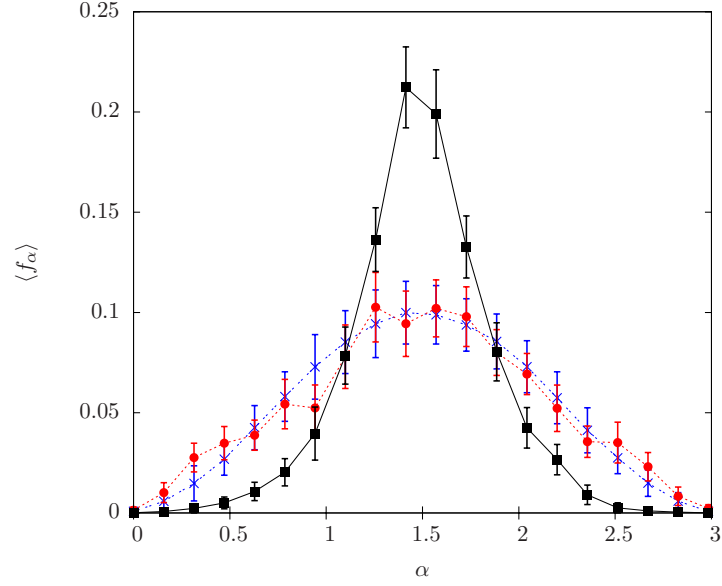}
\end{center}
\caption{The distribution of the average color angle $\left\langle \protect%
\alpha \right\rangle $ between nearest neighbor pseudoparticles of equal
(black squares, full line) and opposite (red bullets, dotted line)
topological charge at $g^{2}=1$. For comparison, we also show $\left\langle 
\protect\alpha \right\rangle $ for a random pseudoparticle distribution
(i.e. at $g^{2}=\infty $; blue crosses, dashed line) which is independent of
the relative topological charge of the neighbors. (The error bars indicate
the statistical standard deviation of the averages per bin.)}
\label{colangfig}
\end{figure}

The resulting $\left\langle \alpha \right\rangle $ distributions are divided
into 20 bins and plotted in Fig. \ref{colangfig} for both equally and
oppositely charged neighbors. For comparison, we also show the angle
distribution of the non-interacting $g^{2}=\infty $ random ensemble. All
three distributions are relatively broadly peaked at $\left\langle \alpha
\right\rangle =\pi /2$. This happens even in the random ensemble, which
indicates that this peak position is statistically favored. Indeed, $\alpha
=\pi /2$ corresponds to the equator of $S^{3}$ and thus to the $\alpha $
value with the maximal number of color orientations. The main lesson of Fig. %
\ref{colangfig} is, however, that equally and oppositely charged dimeron
neighbors on average prefer remarkably different relative color
orientations. While the distribution of the mean color angle $\left\langle
\alpha _{D\bar{D}}\right\rangle $ between nearest neighbors of opposite $Q$
is essentially consistent with a random distribution, the peak value of the $%
\left\langle \alpha _{DD,\bar{D}\bar{D}}\right\rangle $ distribution for
neighbors of equal $Q$ is dynamically enhanced by about a factor of two.

These results can be rather directly understood by recalling that the
interactions between the strongly contracted dimerons in the rather dilute $%
g^{2}=1$ ensemble (with $\left\langle d\right\rangle >\rho $, cf. Fig. \ref%
{ndistfig}) are similar to those among instantons in singular-gauge
instanton-antiinstanton superpositions \cite{sch98,dia84}. Between
instantons of equal topological charge these classical interactions $\Delta
S_{II,\bar{I}\bar{I}}$ are repulsive for any relative color orientation.
More specifically, at distances $d>\rho $ one has \cite{dia84} 
\begin{equation}
\Delta S_{DD,\bar{D}\bar{D}}\overset{g^{2}=1}{\sim }\Delta S_{II,\bar{I}\bar{%
I}}=\frac{32\pi ^{2}}{g^{2}}\left[ 2+\left( 1-2\sin ^{2}\frac{\alpha }{2}%
\right) ^{2}\right] \frac{\rho ^{6}}{d^{6}}+O\left( \frac{\rho ^{8}}{d^{8}}%
\right) .  \label{sii}
\end{equation}%
Since the average distance between $DD$ and $\bar{D}\bar{D}$ pairs in our $%
g^{2}=1$ ensemble is $\left\langle d_{DD,\bar{D}\bar{D}}\right\rangle \simeq
0.93\sim 2\rho $ (cf. Fig. \ref{ndistfig}), the leading term in Eq. (\ref%
{sii}) should reasonably well approximate $\Delta S_{DD,\bar{D}\bar{D}}$.
The resulting $\alpha $ distribution is thus symmetric around $\pi /2$ and
the individual repulsion is minimal (maximal) at $\alpha =\pi /2$ ($\alpha
=0,\pi $). Hence the average repulsion in the ensemble is reduced by more
strongly populating the $\alpha \simeq \pi /2$ color orientations. This
explains the enhanced peak around $\left\langle \alpha _{DD,\bar{D}\bar{D}%
}\right\rangle =\pi /2$ in Fig. \ref{colangfig}.

The instanton-anti-instanton interactions in singular-gauge instanton
superpositions, on the other hand, contain the for $d>\rho $ leading
dipole-dipole interaction \cite{dia84} 
\begin{equation}
\Delta S_{D\bar{D}}\overset{g^{2}=1}{\sim }\Delta S_{I\bar{I}}=-\frac{32\pi
^{2}}{g^{2}}\left[ 1-4\left( c_{\mu }\hat{d}_{\mu }\right) ^{2}\right] \frac{%
\rho ^{4}}{d^{4}}+O\left( \frac{\rho ^{6}}{d^{6}}\right)  \label{siibar}
\end{equation}%
where the unit vector $\hat{d}_{\mu }$ points from the instanton center to
the anti-instanton center. For spacial and color orientations with $%
\left\vert c_{\mu }\hat{d}_{\mu }\right\vert <1/2$, which are predominant in
the $g^{2}=1$ equilibrium ensemble, the leading term in Eq. (\ref{siibar})
is attractive. In fact, indirect evidence for this attraction was already
deduced from our dimeron distributions in Sec. \ref{topdissec}. All
remaining $I\bar{I}$ interactions are repulsive, on the other hand, as those
between equally charged neighbors. In contrast to the latter case, however,
the impact of $\Delta S_{D\bar{D}}$ on the $\left\langle \alpha _{D\bar{D}%
}\right\rangle $ distribution is less transparent since the strength of the $%
D\bar{D}$ attraction is not determined by $\alpha $. Moreover, the
relatively small average dimeron-antidimeron distance $\left\langle d_{D\bar{%
D}}\right\rangle \simeq 0.7\sim 3\rho /2$ for $g^{2}=1$ (cf. Fig. \ref%
{ndistfig}) indicates that the repulsive terms of $O\left( \rho
^{6}/d^{6}\right) $ remain relevant as well. Therefore it seems likely that
this repulsion, together with the overall short-distance repulsion (cf. Sec. %
\ref{topdissec}) and the ensemble entropy, on average compensates at least
part of the leading attraction.\ The resulting \emph{averaged} $D\bar{D}$
correlations should thus be considerably weaker than those in the $DD$ and $%
\bar{D}\bar{D}$ channels. As a consequence, the $\left\langle \alpha _{D\bar{%
D}}\right\rangle $ distribution will be close to that of a random ensemble,
and this is indeed what we observe in Fig. \ref{colangfig}.

The above reasoning focused on two-pseudoparticle forces. This is legitimate
because many-body interactions in the rather dilute $g^{2}=1$ ensemble are
suppressed by the small packing fraction of the contracted dimerons and by
their reduced overlap. Moreover, the emerging parallels with the color
correlations in instanton-anti-instanton superpositions reinforce our
premise that the properties of weakly coupled dimeron ensembles indeed
approach those of instanton liquid models. As already alluded to, the above
arguments further suggest that the smaller average distance $\left\langle
d_{D\bar{D}}\right\rangle <\left\langle d_{DD,\bar{D}\bar{D}}\right\rangle $
between oppositely charged dimerons and the related screening-cloud\
arrangement in the $g^{2}=1$ ensemble (cf. Sec. \ref{topdissec}) are mainly
caused by attractive color-dipole interactions. With growing coupling and
entropy, however, the relative impact of the potentials (\ref{sii}) and (\ref%
{siibar}) on the free energy and on the ensemble structure decreases. This
holds in particular for the $D\bar{D}$ attraction which has the longest
range. Our finding of $\left\langle d_{D\bar{D}}\right\rangle \sim
\left\langle d_{DD,\bar{D}\bar{D}}\right\rangle $ for $g^{2}\gtrsim 25$ (cf.
Fig. \ref{ndistfig}) indeed indicates that the common repulsion in both $D%
\bar{D}$ and $DD,\bar{D}\bar{D}$ channels and the growing entropy begin to
dominate at larger couplings. One would thus expect the peak in the $%
\left\langle \alpha _{DD,\bar{D}\bar{D}}\right\rangle $ distribution to
broaden with increasing coupling until the random distribution is reached
for $g^{2}\rightarrow \infty $. The $\left\langle \alpha _{D\bar{D}%
}\right\rangle $ distribution, on the other hand, may become somewhat more
sharpely peaked at larger $g^{2}$ when the compensating impact of the
attraction subsides.

In order to test these expectations and to shed further light on the
coupling dependence of the color correlations, we have evaluated the
standard deviation $\sqrt{\Delta \left\langle \alpha \right\rangle }$ of the 
$\left\langle \alpha \right\rangle $ distribution from its ($g$ independent)
mean value $\left\langle \alpha \right\rangle =\pi /2$. In Fig. \ref%
{stderfig} we plot $\sqrt{\Delta \left\langle \alpha \right\rangle }$, again
separately for equally and oppositely charged neighbor dimerons, at five $%
g^{2}$ values. For $g^{2}=1$ the standard deviation of the $\left\langle
\alpha _{DD,\bar{D}\bar{D}}\right\rangle $ distribution reaches only about
half of that of the $\left\langle \alpha _{D\bar{D}}\right\rangle $
distribution, in agreement with its higher and narrower peak in Fig. \ref%
{colangfig}. Also expected from Fig. \ref{colangfig} is that the $g^{2}=1$
value of $\sqrt{\Delta \left\langle \alpha _{D\bar{D}}\right\rangle }$
remains close to the random ensemble value at $g^{2}=\infty $. In fact, the
width of the $\left\langle \alpha _{D\bar{D}}\right\rangle $ distribution
decreases only little with growing coupling, as anticipated above, until $%
\left\langle \alpha _{D\bar{D}}\right\rangle $ becomes randomly distributed.
The width of the $\left\langle \alpha _{DD,\bar{D}\bar{D}}\right\rangle $
distribution, on the other hand, grows rather strongly with $g^{2}$ on its
approach to the random value, again confirming our above expectations.
Nevertheless, even at $g^{2}=1000$ the $\left\langle \alpha _{DD,\bar{D}\bar{%
D}}\right\rangle $ distribution remains significantly more peaked than the
random distribution. Hence some of the stronger average repulsion between
nearest-neighbor dimerons of equal topological charge seems to remain
influential even at these rather large couplings (see also Sec. \ref%
{topdissec}).

Reflecting upon the above analysis one may wonder to what extent pertinent
features of the $\alpha $ distributions are obscured by taking the
configuration and ensemble averages. Indeed, the spacial averaging and the
interplay between spacial and color correlations may wash out interesting
local features of the relative color ordering. However, one would expect
this type of leveling to be weakest in the $g^{2}=1$ ensemble where the
almost spherical color distribution of the dimerons is most concentrated and
the impact of their spacetime orientation consequently minimized. Other
qualitative effects of the stronger intermediate-range order among
neighboring dimerons at $g^{2}=1$ may therefore also be robust enough to
survive the averaging procedure.

\subsection{Topological susceptibility}

\label{tssec}

The topological susceptibility $\chi _{t}$ characterizes several fundamental
properties of the Yang-Mills vacuum. It governs, for instance, the
dependence of the free energy on the CP violating vacuum angle $\theta $
around $\theta =0$ and the mass of the $\eta ^{\prime }$ meson in quantum
chromodynamics with a large number $N_{c}$ of colors \cite{wv}. Moreover, it
contains information on the interplay between the Yang-Mills dynamics and
the topology of the gauge group which was conjectured long ago to be the
origin of linear quark confinement \cite{pol77}.

In a finite spacetime volume $V$, the topological susceptibility is defined
as 
\begin{equation}
\chi \left( V\right) =\frac{\left\langle Q\left( V\right) Q\left( V\right)
\right\rangle }{V}  \label{chitop}
\end{equation}%
(which may contain contact terms \cite{tsren} in the $\rho \rightarrow 0$
and $a\rightarrow 0$ limits) where $Q\left( V\right) $ is the topological
charge of the gauge field in $V$, i.e. 
\begin{equation}
Q\left( V\right) =\int_{V}d^{4}xq\left( x\right) \text{ \ \ \ \ \ with }%
q\left( x\right) =\frac{1}{16\pi ^{2}}tr\left\{ F_{\mu \nu }\widetilde{F}%
_{\mu \nu }\right\}  \label{q}
\end{equation}%
($\widetilde{F}_{\mu \nu }\equiv \varepsilon _{\mu \nu \alpha \beta
}F_{\alpha \beta }/2$). In the context of our vacuum description by
topologically active constituents, $\chi $ is of particular interest because
it quantifies the strength of topological charge fluctuations around $%
\left\langle Q\right\rangle =0$. The latter is enforced by CP symmetry which
is sufficiently manifest in our ensembles because the configurations (\ref%
{cf}) contain a very nearly equal number of dimerons and anti-dimerons 
\footnote{%
Even for $N_{D}=N_{\overline{D}}$ strict CP invariance would hold only in
the infinite-volume limit and/or with infinite ensemble statistics,
incidentally, since the topological charge density of the dimerons is not
fully contained in our evaluation volumes. However, this boundary artefact
is mitigated already in relatively small volumes by the short-range
anticorrelations among the dimerons' topological charges (cf. Sec. \ref%
{dimenssec}).}.

The topological susceptibility $\chi _{t}\equiv \lim_{V\rightarrow \infty
}\chi \left( V\right) $ of SU$\left( 2\right) $ Yang-Mills theory is known
from several independent high-precision lattice simulations \cite%
{tslatt,luc01} (with their scale set by prescribing the value of the
physical string tension). At large $N_{c}$, furthermore, the value of $\chi
_{t}$ can be obtained either from lattice extrapolations \cite%
{luc01,tep02,luc04} or from experimentally measured properties of the
lightest pseudoscalar meson nonet. Indeed, to leading order in $1/N_{c}$ the
Witten--Veneziano mechanism predicts \cite{wv}%
\begin{equation}
\chi _{t,\text{SU}\left( N_{c}\rightarrow \infty \right) }\simeq \frac{%
f_{\pi }^{2}}{2N_{f}}\left( m_{\eta ^{\prime }}^{2}+m_{\eta
}^{2}-2m_{K}^{2}\right) \simeq \left( 180\text{ MeV}\right) ^{4}
\label{wvrel}
\end{equation}%
(where $f_{\pi }$ is the pion decay constant, $N_{f}$ is the number of
flavors and $m_{\eta },m_{\eta ^{\prime }},m_{K}$ are the $\eta $, $\eta
^{\prime }$ and $K$ meson masses) and thereby relates $\chi _{t,\text{SU}%
\left( \infty \right) }$ to the part of the $\eta ^{\prime }$ mass which
does not originate from the strange-quark mass. The above values of $\chi
_{t}$ will serve as benchmarks for comparison with our results and for
scale-setting purposes.

To evaluate $\chi $ in our dimeron ensembles, we first compute the
topological charge $Q\left( V\right) $ of each gauge-field configuration.
More specifically, we integrate the topological charge density, obtained
analytically from Eqs. (\ref{cf}) and (\ref{q}), on a special grid \footnote{%
Our $V$ resolution is obtained by using volumes with extent $1.2$ and step
size $0.1$ in the three spacial directions, and a step width of $0.03$ in
the temporal direction which maximally extends to $4.0$.} whose extent is
varied in small steps to tune through the desired range of volumina $V.$ The
correlation between discretization errors in the two $Q$ factors of $\chi $
(which arise from neglecting fluctuations with wavelengths below the
grid-point distance in the coarse-grained density $q$) is reduced by
evaluating the second factor on a modified grid. The latter is obtained by
replacing each of the original grid-point positions with a randomly chosen
one inside a surrounding volume determined by the inverse grid-point
density. Finally, the ensemble average is carried out according to Eq. (\ref%
{chitop}).

Translational invariance implies that $\chi \left( V\right) $ becomes
volume-independent in the thermodynamic limit and that $\chi _{t}\equiv
\lim_{V\rightarrow \infty }\chi \left( V\right) $ exists (after appropriate
renormalization). In fact, this $V$ independence will develop already in
finite volumes whose linear dimensions $\sim V^{1/4}$ sufficiently much
exceed the correlation length of the topological charge density. In order to
test how far this applies to our simulation volumes and permits
extrapolations of our $\chi \left( V\right) $ to $V\rightarrow \infty $, we
have calculated $\chi \left( V\right) $ in a range of volumes in which
boundary effects remain controllable. (Numerical efficiency is improved by
varying only the temporal extent of the evaluation volumes. All three
spacial dimensions of the boxes are kept fixed at $1.2$ (chosen to limit
boundary artefacts, cf. Sec. \ref{eran}) and remain centrally embedded.)

\begin{figure}[tbp]
\begin{center}
\includegraphics[height = 9cm]{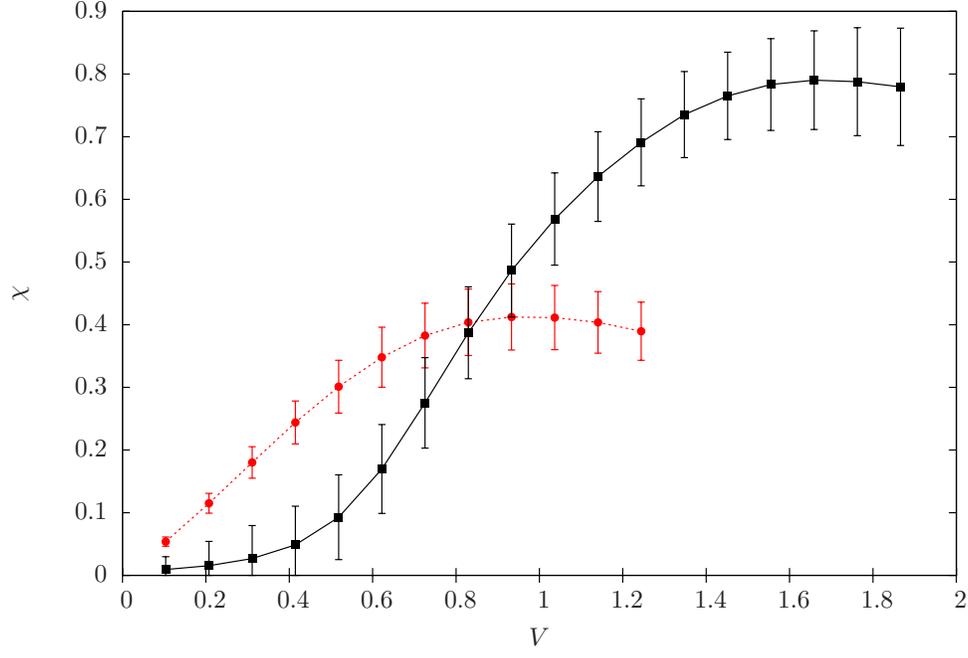}
\end{center}
\caption{The topological susceptibility $\protect\chi $ as a function of the
spacetime volume $V$ for $g^{2}=1$ (red bullets, dotted line) and $g^{2}=100$
(black squares, full line).}
\label{ts}
\end{figure}

The resulting volume dependence of the topological susceptibility is shown
in Fig. \ref{ts} for $g^{2}=1$ and $100$. In small volumes topological
charge fluctuations are suppressed and $\chi \left( V\right) $ accordingly
starts out close to zero. Towards our largest reliably accessible volumes,
on the other hand, boundary effects may become relevant and begin to reduce $%
\chi $. (Hence Fig. \ref{ts} contains independent information on the extent
to which the boundary breaks translational invariance.) Nevertheless, for
both coupling values $\chi \left( V\right) $ shows a rather broad maximum
towards the end of the trustworthy $V$ range. It is tempting to interpret
these maxima as the onset of the expected saturation plateaus, and the
corresponding $\chi $ values as reliable approximations to the
infinite-volume predictions $\chi _{t}$. However, from our data sets one
cannot decide with confidence whether the maxima do not instead
underestimate $\chi _{t}$ by interpolating between the rise of $\chi \left(
V\right) $ at small $V$ and its boundary-induced decline at large $V$. We
shall therefore interpret the maximal values $\hat{\chi}$ (in this and the
following sections a hat will indicate a dimensionful quantity given in our
numerical units) more conservatively as lower bounds on the topological
susceptibility.

%TCIMACRO{\TeXButton{B}{\begin{table}[htbp] \centering}}%
%BeginExpansion
\begin{table}[htbp] \centering%
%EndExpansion
\begin{tabular}{|c|c|c|c|}
\hline
$g^{2}$ & $\hat{\chi}^{1/4}$ & $a_{\chi }$ [fm] & $\sigma $ [fm$^{-2}$] \\ 
\hline\hline
\multicolumn{1}{|l|}{$1$ \ \ \ \ \ \ } & $0.75\pm 0.032$ & $0.076\pm 0.0032$
& $3.\,\allowbreak 864\pm 0.0793$ \\ \hline
\multicolumn{1}{|l|}{$25$} & $0.83\pm 0.043$ & $0.084\pm 0.0045$ & $9.832\pm
0.7271$ \\ \hline
\multicolumn{1}{|l|}{$100$} & $0.94\pm 0.032$ & $0.095\pm 0.0033$ & $%
9.340\pm 0.4518$ \\ \hline
\multicolumn{1}{|l|}{$1000$} & $2.42\pm 0.083$ & $0.246\pm 0.0096$ & $%
3.363\pm 0.1051$ \\ \hline
\multicolumn{1}{|l|}{$\infty $} & $10.35\pm 0.503$ & $1.043\pm 0.0508$ & $%
0.1607\pm 0.0008514$ \\ \hline
\end{tabular}%
\caption{(Lower bounds on) the fourth root of the topological susceptibility $\hat \chi$ in numerical 
units for five values of the squared gauge coupling. The last columns contain the 
grid constant $a_\chi$ and the string tension $\sigma$ in physical units as 
obtained when setting the scale with the SU(2) Yang-Mills value 
$\chi^{1/4}=0.99 \, {\rm fm}^{-1}$.
\label{tstab}} 
%TCIMACRO{\TeXButton{E}{\end{table}}}%
%BeginExpansion
\end{table}%
%EndExpansion

In Table \ref{tstab} the corresponding values of $\hat{\chi}^{1/4}$\textbf{\ 
}are listed with statistical (jackknife) error estimates for our five
standard $g^{2}$ values. On a cautionary note, we recall that the $g^{2}=1$
dimeron ensemble thermalizes very slowly and that the autocorrelation for
the topological charge becomes correspondingly large. The error of $\hat{\chi%
}^{1/4}$ as quoted in Table \ref{tstab} does not take this strong
autocorrelation into account and is therefore probably underestimated at
least for $g^{2}=1$. With this caveat in mind, we observe that $\hat{\chi}%
^{1/4}$ stays inside our error estimates for $25\leq g^{2}\leq 100$
practically constant (which becomes even more obvious when expressing $\chi
^{1/4}$ in physical units, cf. Table \ref{respu}). This weak coupling
dependence may be related to the similarly weak temperature dependence of
the topological susceptibility in the confined phase of Yang-Mills theory up
to the phase transition \cite{luc04}. For larger $g^{2}$ our $\hat{\chi}%
^{1/4}$ values grow too strongly, on the other hand, which is at least in
part due to boundary artefacts. Indeed, from $g^{2}\gtrsim 1000$ the
dimerons are on average as far dissociated as the boundary allows. As
discussed in Sec. \ref{eran}, they then tend to accumulate along the box
diagonals, with their centers concentrating in the middle of the box. These
distorted field configurations break translational invariance almost
everywhere in the simulation volume.

Our numerical results for the topological susceptibility provide a first
opportunity to set the distance scale $a$ in our box as a function of $g^{2}$%
. To this end we write $a_{\chi }=\hat{a}\left( \hat{\chi}/\chi \right)
^{1/4}$ in physical units, where $\hat{a}=0.1$ and $\hat{\chi}$ are given in
our numerical units while $\chi =\left( 0.99\right) ^{4}$ fm$^{-4}$ is the SU%
$\left( 2\right) $ Yang-Mills value obtained from the SU$(2)$ lattice result 
$\chi ^{1/4}/\sigma ^{1/2}=0.483\pm 0.006$ \cite{luc01} in combination with
the physical string tension $\sigma \simeq 4.2$ fm$^{-2}$ \cite{mil}. The
resulting values of $a_{\chi }$ in fm are listed in Table \ref{tstab}. With
their help one may convert our data e.g. for the string tension $\hat{\sigma}
$ (cf. Table \ref{wltab}) via $\sigma =\left( \hat{a}/a_{\chi }\right) ^{2}%
\hat{\sigma}$ into physical units, as done in the last column of Table \ref%
{tstab}.

A quantitative comparison of our results with those of other approaches in
physical units will be postponed to Sec. \ref{sindep}. For a first
orientation, however, one may compare our \textquotedblleft raw
data\textquotedblright\ for $\hat{\chi}$ with their counterparts from the
regular-gauge instanton and meron ensembles of Ref. \cite{len08} (which
adopted $g^{2}=32$ and a size regulator $\rho \leq 0.1$) even though the
physical distance scales might not be straightforwardly related. (In the
ensembles of Ref. \cite{len08} about 90\% of the topological susceptibility
is generated by the peaks in the topological charge density associated with
individual merons or regular-gauge instantons, incidentally; the remaining,
smoother background field contributes just the remaining 10\%.) According to
Table \ref{tstab}, our ensembles with $g^{2}=25-100$ give $\hat{\chi}%
^{1/4}\sim 0.8-0.9$ which is of the same order as the meron-ensemble result $%
\hat{\chi}^{1/4}=0.77$ \cite{len08} for a comparable meron density in
physical units,\textbf{\ }corresponding to $N_{M+\bar{M}}=100$ (see below).
In fact, at the above couplings $g^{2}=25$ and $100$ our dimeron density
reaches about half of Ref. \cite{len08}'s meron density and therefore yields
a similar density of meron constituents. (The regular-gauge instanton
ensemble with an instanton density roughly equal to our dimeron density
(corresponding to $N_{I+\bar{I}}=100$), on the other hand, generates the
significantly larger value $\hat{\chi}^{1/4}\simeq 1.37$ \cite{len08}.)
Within errors the\ topological susceptibilities of meron and dimeron
ensembles in numerical units are therefore consistent. As a consequence,
setting scales by imposing a common \textquotedblleft
physical\textquotedblright\ value for $\chi \ $will lead to comparable
physical\ distance scales.

\subsection{Wilson loops, static-quark potentials and string tension}

\label{wlsec}

The search for the conjectured transition of dimeron ensembles into
confining meron ensembles is one of the central objectives of our study (cf.
Sec. \ref{introsec}). Our first quantitative evidence for the actual
development of a meron-dominated phase was discussed in Sec. \ref{dimenssec}
where we found the dimerons to dissociate into their progressively
independent meron constituents when the coupling increases. We are now
proceeding to the crucial question whether the gradually liberated merons
indeed generate an area law for large Wilson loops and thus linear
confinement with a finite string tension. More specifically, we are going to
evaluate ensemble averages (i.e. vacuum expectation values) of path-ordered
Wilson loops%
\begin{equation}
W\left[ C\left( R,T\right) \right] =tr\left( \mathcal{P}\exp
i\doint\limits_{C}A_{\mu }dx_{\mu }\right)  \label{wl}
\end{equation}%
along closed, rectangular paths $C$. The latter correspond to the worldlines
of fundamentally colored, static quark and antiquark sources, separated by a
spacial distance $R$ and evolving over the Euclidean time $T>R$. The static
energy required to introduce and separate these sources in the vacuum is
then given by the potential%
\begin{equation}
V\left( R\right) =-\lim_{T\rightarrow \infty }\frac{1}{T}\ln \left\langle W%
\left[ C\left( R,T\right) \right] \right\rangle .  \label{V(R)def}
\end{equation}%
Hence a linear rise of $V\left( R\right) $ for sufficiently large $R$
signals linear quark confinement by color-flux tube formation.

Since we are particularly interested in the behavior of $V\left( R\right) $
at the largest reliably calculable $R$ values, our strategy for computing $W$
is designed to minimize the impact of boundary effects (as identified in
Sec. \ref{eran}) while still making efficient use of the ensemble
configurations. This is achieved by restricting the evaluation volume to
several subvolumina centered in the core box of Sec. \ref{box}. Artefacts of
the boundary are reduced by limiting the extent of these volumina to $2.0$
in the time direction, to $1.0$ in the spacial $R$ direction and to $0.5$ in
the two remaining spacial directions. The volumina generated by permutations
of the spacial directions are included as well. As a consequence, fields in
regions in which we diagnosed the main violations of translational
invariance are excluded from the evaluation.

Next, we compute the oriented links $U\left[ l\left( x\rightarrow x\right) %
\right] =tr\mathcal{P}\exp i\int_{l\left( x\rightarrow x^{\prime }\right)
}A_{\rho }dx_{\rho }$ which connect neighboring grid points $x_{\mu }$ and $%
x_{\nu }^{\prime }$ in the evaluation volume by straight lines $l\left(
x\rightarrow x^{\prime }\right) $. All rectangular Wilson loops with aspect
ratio $R/T=1/2$ are then constructed from combinations of these color
parallel transporters. (For small enough loops we exploit Euclidean symmetry
to assign the time direction to the longer side, irrespectively of its
orientation in the core box.) In order to obtain quantitative information
about the impact of boundary and resolution errors on the reliability of the
link calculation, we perform the latter both (i) on the full subvolumina
described above, with the $R$ values progressing in steps of length $0.1$,
and (ii) just on the central hyperplanes of these volumina, spanned by the
longer spacial direction (with length $1.0$) and the temporal direction, in
which the $R$ values progress with step size $0.025$.

The main challenge for our finite grid resolution arises from gauge fields
along a link which develop unusually large gradients by approaching a meron
center, as already alluded to in Sec. \ref{eran}. Since this turns out to
happen relatively rarely, we decided to dynamically adapt the resolution of
the grid on which the gauge fields are sampled along the link. The refined
grid then increases the accuracy of the numerical integration in the link
exponent. We start with a maximal grid-point distance of $0.01$, i.e. on a
grid which is 10 times finer than the one on which the gauge-field
configurations were generated. This distance is reduced by half if the
Frobenius norm of the difference between a given link and the path-ordered
product of two links over half the distance exceeds one per mille, i.e. if 
\begin{equation}
\left\Vert U\left[ l\left( x_{\mu }\rightarrow x_{\mu +1}\right) \right] -U%
\left[ l\left( x_{\mu }\rightarrow x_{\mu +1/2}\right) \right] U\left[
l\left( x_{\mu +1/2}\rightarrow x_{\mu +1}\right) \right] \right\Vert \geq
0.001.  \label{crit}
\end{equation}%
The criterion (\ref{crit}) is chosen such that about one third of the links
in an average Wilson loop has to be updated at least once. (Two refinement
updates of the subgrid for the same link are almost never encountered with
our regulator $\rho ^{2}=0.3$.) Finally, we calculate the average over all
Wilson loops \footnote{%
The values of very small Wilson loops\ could alternatively be obtained
analytically since our gauge field configurations (\ref{cf}) become constant
at length scales $\ll {\protect \rho}$ (cf. Ref. \cite{len08}).} 
with the same edge
length $R$ and evaluate the logarithm $\ln \left\langle W\left[ C\left(
R,2R\right) \right] \right\rangle $ of their ensemble average.

\begin{figure}[tbp]
\begin{center}
\includegraphics[height = 9cm]{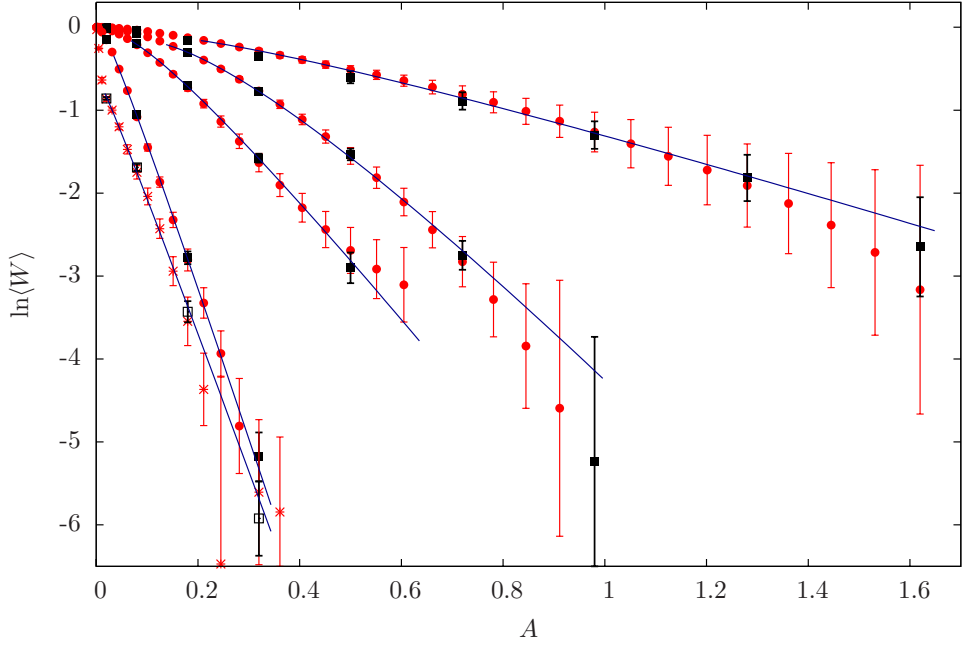}
\end{center}
\caption{The logarithm of the expectation value of rectangular Wilson loops
with area $A$ (and side length $R=T/2=\protect\sqrt{A/2})$ for the square
coupling values $g^{2}=1,$ $25,$ $100,$ $1000$ and $\infty $. (For
increasing coupling the data lie below each other.) The black squares
(boxes) are obtained from the loops in the full evaluation volume, the red
dots (asterisks) from the loops in the central hyperplanes. The fit curves
to $f\left( A\right) $, as described in the text, are drawn as full lines in
their fit intervals. }
\label{wlfig}
\end{figure}

In Fig. \ref{wlfig} the resulting data are plotted as a function of the loop
area $A=2R^{2}$ for our five square couplings $g^{2}=1$, $25$, $100,$ $1000$
and $\infty $. The figure includes the points obtained from both the full
subvolumina and their central hyperplanes, which agree within errors 
\footnote{%
The statistical error of the data points from the subvolumina is somewhat
smaller than that from their central hyperplanes. When plotting $%
\left\langle W\left[ C\left( R,2R\right) \right] \right\rangle $ as a
function of $R$, incidentally, some seemingly correlated discrepancies
between these data sets become nevertheless visible for $g^{2}=1$ but not
for larger couplings. Hence these distortions may indeed originate from
boundary effects (cf. Sec. \ref{eran}) and appear to be sufficiently
suppressed in the chosen subvolumina.}. It further shows the best fits of
the $\ln \left\langle W\right\rangle $ data to the function%
\begin{equation}
f\left( A\right) =\hat{\omega}+\hat{\tau}P\left( A\right) -\hat{\sigma}A
\label{f}
\end{equation}%
where $P\left( A\right) =3\sqrt{2A}$ is the perimeter of the rectangular
loops. The ansatz (\ref{f}) may e.g. be motivated by a string model which
takes small, UV-regularized quantum fluctuations into account \cite%
{die83,len08}. All fit results are collected in Table \ref{wltab}, together
with their $\chi ^{2}$\ values per degree of freedom and the fit ranges $%
A\in \left[ A_{\text{min}},A_{\text{max}}\right] $ in which the data are
probably reliable. The $\chi ^{2}$\ values indicate that the fits indeed
allow for a reliable determination of the parameters $\hat{\omega},$ $\hat{%
\tau}$ and in particular of the string tension $\hat{\sigma}$. The central
lesson of these results is that dimeron ensembles indeed generate finite
string tensions which monotonically increase with $g^{2}$. Our findings in
Secs. \ref{dissocsec} -- \ref{colcorsec} explain this increase as a
consequence of the dimerons' gradual release of their disordering meron
tails. In addition, our results provide evidence for dimeron dissociation to
trigger the transition from a non-confining regime to a confining phase. (In
the thermodynamic limit the analogy with Kosterlitz-Thouless-type
transitions suggests that the string tension will vanish exactly below some
critical value $g_{c}^{2}$. In finite volumes one expects a more gradual
transition, however, as seen in our data.)

%TCIMACRO{\TeXButton{B}{\begin{table}[htbp] \centering}}%
%BeginExpansion
\begin{table}[htbp] \centering%
%EndExpansion
\begin{tabular}{|c|c|c|c|c|c|c|}
\hline
$g^{2}$ & $\hat{\omega}$ & $\hat{\tau}$ & $\hat{\sigma}$ & $\chi ^{2}$ & $A_{%
\text{min}}$ & $A_{\text{max}}$ \\ \hline\hline
\multicolumn{1}{|l|}{$1$ \ \ \ \ \ \ } & $-0.22\pm 0.09$ & $0.27\pm 0.07$ & 
\multicolumn{1}{|c|}{$2.23\pm 0.22$} & $0.22$ & $0.20$ & $1.60$ \\ \hline
\multicolumn{1}{|l|}{$25$} & $-0.51\pm 0.04$ & $0.81\pm 0.04$ & 
\multicolumn{1}{|c|}{$6.95\pm 0.21$} & $0.32$ & $0.10$ & $0.90$ \\ \hline
\multicolumn{1}{|l|}{$100$} & $-0.20\pm 0.01$ & $0.55\pm 0.02$ & 
\multicolumn{1}{|c|}{$8.47\pm 0.18$} & $0.44$ & $0.03$ & $0.51$ \\ \hline
\multicolumn{1}{|l|}{$1000$} & $-0.02\pm 0.04$ & $0.47\pm 0.08$ & 
\multicolumn{1}{|c|}{$20.08\pm 0.66$} & $0.21$ & $0.03$ & $0.35$ \\ \hline
\multicolumn{1}{|l|}{$\infty $} & $-0.63\pm 0.09$ & $0.23\pm 0.20$ & 
\multicolumn{1}{|c|}{$17.57\pm 1.80$} & $1.06$ & $0.01$ & $0.35$ \\ \hline
\end{tabular}%
\caption{Results for the parameters $\hat \omega$, $\hat \tau$ and the string tension 
$\hat \sigma$, obtained from the fit of $f(A)$ to $\ln \langle W \rangle (A)$ in the interval
from $A_{\rm{min}}$ to $A_{\rm{max}}$. (The reduced chi-square value $\chi^2$  
characterizes the statistical goodness of the fit and should not be confused with 
the square of the topological susceptibility.)
\label{wltab}} 
%TCIMACRO{\TeXButton{E}{\end{table}}}%
%BeginExpansion
\end{table}%
%EndExpansion

\begin{figure}[tbp]
\begin{center}
\includegraphics[height = 9cm]{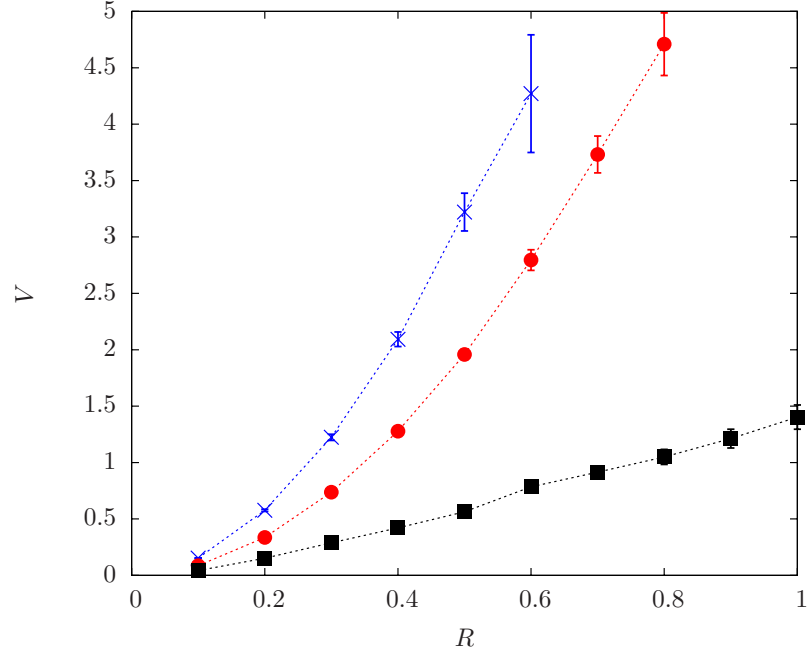}
\end{center}
\caption{The heavy-quark potentials $V(R)$ for the square coupling values $%
g^{2}=1,$ $25$ and $100$. (The stronger potentials correspond to larger
couplings.) }
\label{V(R)}
\end{figure}

We have additionally applied a somewhat complementary method of computing
the string tension. In this approach the static quark-antiquark potential $%
V\left( R\right) $ is extracted directly from the Wilson-loop behavior
according to Eq. (\ref{V(R)def}). This allows us to include loops of
multiple aspect ratios, provides a useful cross check on our results for $%
\hat{\sigma}$ in Table \ref{wltab} and makes it possible to obtain physical
information from the small-$R$ behavior of $V$ as well. To implement this
approach we evaluate rectangular loops of all aspect ratios which fit into
the subvolumina defined above. The value of $V$ at a given $R$ is then
extracted from a fit of $\ln \left\langle W\right\rangle /T$ to a constant
inside the $T\gtrsim R$ ranges in which $\ln \left\langle W\right\rangle /T$
becomes approximately $T$ independent. The resulting potentials are plotted
in Fig. \ref{V(R)} for $g^{2}=1,25$ and $100$. All $V\left( R\right) $
curves indeed show an essentially linear rise for $R\gtrsim 0.4$. (Due to
growing boundary effects, the potentials cannot be reliably extracted beyond
a coupling-dependent $R_{\text{max}}\gtrsim 0.7-1$.) It is reassuring,
furthermore, that fits to the linear region yield values for the string
tension ($\hat{\sigma}\left( g^{2}=1\right) \simeq 1.6,$ $\hat{\sigma}\left(
g^{2}=25\right) \simeq 6.3$ and $\sigma \left( g^{2}=100\right) \simeq 9.3$)
which are within 10 -- 20\% of those with better statistics given in Table %
\ref{wltab}.

Before setting scales and comparing to results of other approaches in Sec. %
\ref{sindep}, we may get a first idea of the quantitative significance of
our $\hat{\omega},\hat{\tau}$ and especially $\hat{\sigma}$ values by
comparing them to their counterparts in the meron and regular-gauge
instanton ensembles of Ref. \cite{len08} (which adopted $g^{2}=32,$ $\rho
\leq 0.1$ and $N_{I+\bar{I}\text{,}M+\bar{M}}=500$). The meron ensemble
yields $\hat{\omega}=-0.5$, $\hat{\tau}=0.94$ and $\hat{\sigma}=12.8$ which
are overall closest to our $g^{2}=25$ and $100$ results. However, the string
tension in the meron ensemble exceeds ours in the $g^{2}=100$ dimeron
ensemble by about $50\%$. This is probably because our dimerons' constituent
merons are not yet fully liberated at $g^{2}=25-100$ (cf. Table \ref%
{dissoctab}) and because their density is somewhat lower than the meron
density in the $N_{M+\bar{M}}=500$ ensemble of Ref. \cite{len08}. In any
case, the \textquotedblleft raw\textquotedblright\ data suggest that the
confinement properties of $g^{2}=32$ meron and $g^{2}=25-100$ dimeron
ensembles are at least qualitatively compatible. (Another indication may be
that the string tensions in both meron and dimeron ensembles roughly double
in the strongly-coupling limit, boundary effects notwithstanding. The
results $\omega =-0.1$, $\tau =0.58$, $\sigma =20.5$ for the regular-gauge
instanton ensemble \cite{len08}, incidentally, turn out to reproduce ours
from the $g^{2}=1000$ dimeron ensemble up to a few percent.)

The above considerations indicate that fixing the string tension at a
common\ physical value will result in comparable\ confinement scales of both
meron and dimeron ensembles in the $g^{2}=25-100$\ coupling region. This is
reassuring because the linear rise of our $V\left( R\right) $ should then
rather directly match on to the linear potential of the\ meron ensemble even
at $R\gtrsim 1$, where the confining flux tube is expected to develop fully
but where we cannot reliably extract it from our present data set. This
expectation is supported by fits of the meron-ensemble result for $V\left(
R\right) $ to a regularized string model in an extended $R$ range including $%
R>1$ \cite{len08}. In fact, within errors those fits reproduce the string
tension obtained from fitting $\ln \left\langle W\right\rangle $ to Eq. (\ref%
{f}) in an $A$ range with a smaller $R_{\text{max}}=\sqrt{A/2}$ than ours.
This indicates that our string tension can indeed be reliably determined by
fitting Eq. (\ref{f}) in a region of relatively small areas. It also adds
considerably to the evidence for dimeron ensembles to provide a viable
pathway towards confining meron ensembles.

We close this section with a few comments on the small-$R$ behavior of $%
V\left( R\right) $ which is of physical interest in its own right. For $%
R\lesssim 0.4$ our potentials exhibit to a good approximation a quadratic $R$
dependence which extends farthest for $g^{2}=1$ and is little affected by
boundary artefacts. In order to understand this behavior one should recall
that such a quadratic small-$R$ potential \footnote{%
Linear short-distance contributions to the static quark potential
(originating e.g. from a nonlocal two-dimensional gluon condensate \cite%
{gub01} and modeled by a tachyonic UV\ gluon mass) were argued for as well 
\cite{akh98}. (It is conceivable that the latter may also contribute in
dimeron ensembles where the linear behavior of $V\left( R\right) $ sets in
at rather small distances.)} is generated by sufficiently dilute
instanton-antiinstanton ensembles \cite{cal78,cal378,dia89}. This indicates
that for $R\lesssim \rho $ our $V\left( R\right) $ are dominated by
contributions from little dissociated dimerons which behave essentially as
singular-gauge instantons. (Table \ref{dissoctab} and the dimeron density
estimates in Sec. \ref{sindep} suggest that our ensembles keep a substantial
fraction of such instanton-like dimerons even at larger couplings.) For a
more quantitative consistency check we recall that a dilute SU$\left(
2\right) $ instanton ensemble generates the short-distance behavior $V\left(
R\right) \simeq 0.5\rho ^{-3}R^{2}$ \cite{cal78}, i.e. $V\left( R\right)
\simeq 2.8$ $R^{2}$ for our size parameter $\rho =0.55$. From a quadratic
fit to our $g^{2}=1$ potential at distances $R\lesssim 0.4$, on the other
hand, we obtain $V\left( R\right) \simeq 3.8$ $R^{2}$ which is indeed of the
same order.

\subsection{Dimensionless amplitude ratios and results in physical units}

\label{sindep}

A quantitative comparison of the dimensionful amplitudes and observables
calculated above with those of other approaches requires either to form
dimensionless combinations of our results or to rewrite them in physical
units. Both will be done in the present section. We start with the
discussion of three dimensionless ratios which can be formed from our
results and thus compared to other data without scale-setting ambiguities.

Our first example is the ratio $\left\langle \hat{s}\right\rangle /\hat{%
\sigma}^{2}=\left\langle s\right\rangle /\sigma ^{2}$ of the
ensemble-averaged Yang-Mills action density $\left\langle \hat{s}%
\right\rangle $ (the hat again indicates numerical units) and the
appropriate power of the string tension given in Table \ref{wltab}. The
density $g^{2}\left\langle \hat{s}\right\rangle $ is (modulo renormalization
issues) proportional to the gluon condensate $\left\langle
F^{2}\right\rangle $ and therefore of considerable interest in its own
right.\ Our results for $g^{2}\left\langle \hat{s}\right\rangle $ can be
obtained from the values computed in Sec. \ref{eran} on diagonals through
the simulation volume\ by taking the average values of the plateau fits in
Figs. \ref{sTg1fig} and \ref{sDg5fig}, i.e. $g^{2}\left\langle
s\right\rangle \left( g^{2}=1\right) \simeq 210$ and $g^{2}\left\langle
s\right\rangle \left( g^{2}=25\right) \simeq 425$. With the results for $%
\hat{\sigma}$\textbf{\ }in Table \ref{wltab} this yields $\left\langle
s\right\rangle /\sigma ^{2}\left( g^{2}=1\right) \simeq 94$ and $%
\left\langle s\right\rangle /\sigma ^{2}\left( g^{2}=25\right) \simeq 3$.
The latter value corresponds to the coupling range $g^{2}\simeq 25-100$ in
which the dimeron ensembles were estimated above to best approximate the
confining phase of the Yang-Mills vacuum. Hence this value should be
compared to the SU$\left( 2\right) $ lattice value $\left\langle
s\right\rangle /\sigma ^{2}\simeq 4.5$ \cite{cam84}, the QCD sum-rule values 
$\left\langle s\right\rangle /\sigma ^{2}\simeq 4-10$ \cite{for205} (which
correspond to $N_{c}=3$ and contain quark corrections, however) and the
value $\left\langle s\right\rangle /\sigma ^{2}\simeq 8$ \footnote{%
In Refs. \cite{len04,len08} it was argued that the meron peak contributions,
which diverge logarithmically for ${\protect \rho} \rightarrow 0$, should be
subtracted from $\left\langle s\right\rangle $ in meron ensembles. The
remaining, about 25-35\% reduced \textquotedblleft
background\textquotedblright\ action density is then considered as the
physical result, renormalized at ${\protect \rho} ^{-1}$. For our more localized
singular-gauge dimerons this procedure will be less effective, and we shall
not attempt to implement it here.} in the meron ensemble \cite{len08}. (The
regular-gauge instanton ensemble yields $\left\langle s\right\rangle /\sigma
^{2}\simeq 10$ \cite{len08}.). Of course, all these values should be taken
with a grain of salt since they involve large subtractions whose systematic
error cannot be reliably estimated. In any case, within expected errors our
result for $g^{2}=25$ is compatible with the SU$\left( 2\right) $ lattice
value and several QCD sum-rule estimates.

A second useful dimensionless ratio to be assembled from our results is $%
\chi /\left( g^{2}\left\langle s\right\rangle \right) $. In fact, there has
been a proposal for an approximate low-energy relation in Yang-Mills theory, 
$\chi \simeq \left\langle F^{2}\right\rangle /\left( 66\pi ^{2}N_{c}\right) $
\cite{hal98}, which would fix this ratio as $\chi /\left( g^{2}\left\langle
s\right\rangle \right) \simeq 1/\left( 33\pi ^{2}\right) \simeq 3.07\times
10^{-3}$ (for $N_{c}=2$ and with $g^{2}\left\langle s\right\rangle
=\left\langle F^{2}\right\rangle /4$). A dilute-instanton-gas estimate
similarly yields $\chi /\left( g^{2}\left\langle s\right\rangle \right)
\simeq 1/\left( 32\pi ^{2}\right) $ \footnote{%
The relation given in Ref. \cite{sch98} contains an error. We thank Thomas
Sch\"{a}fer for clarifying correspondence on this issue.}. With the SU$%
\left( 2\right) $ Yang-Mills lattice result $\chi ^{1/4}/\sigma
^{1/2}=0.483\pm 0.006$ \cite{luc01} and the ratio quoted above one arrives
at the similar value $\chi /\left( g^{2}\left\langle s\right\rangle \right)
\simeq 3.21\times 10^{-3}$ while the range of condensate results $%
\left\langle F^{2}\right\rangle \simeq 0.1-0.3$ GeV$^{4}$ \cite{for205} from
QCD sum-rule analyses (corresponding to $N_{c}=3$ and containing quark
admixtures)\ together with the SU$\left( 2\right) $ Yang-Mills value $\chi
^{1/4}\simeq 195~$MeV (see below) yields $\chi /\left( g^{2}\left\langle
s\right\rangle \right) \simeq \left( 1.2-3.6\right) \times 10^{-3}$. From
the data for $\hat{\chi}$ in Table \ref{tstab} and for $\left\langle \hat{s}%
\right\rangle $\ as given above, finally, our dimeron ensemble results are $%
\chi /\left( g^{2}\left\langle s\right\rangle \right) \simeq 1.9\times
10^{-3}$ for $g^{2}=1$ and $\chi /\left( g^{2}\left\langle s\right\rangle
\right) \simeq 1.\,2\times 10^{-3}$ for $g^{2}=25$, within the range
obtained from the sum-rule values for the gluon condensate but smaller than
the lattice value.

\begin{figure}[tbp]
\begin{center}
\includegraphics[height = 9cm]{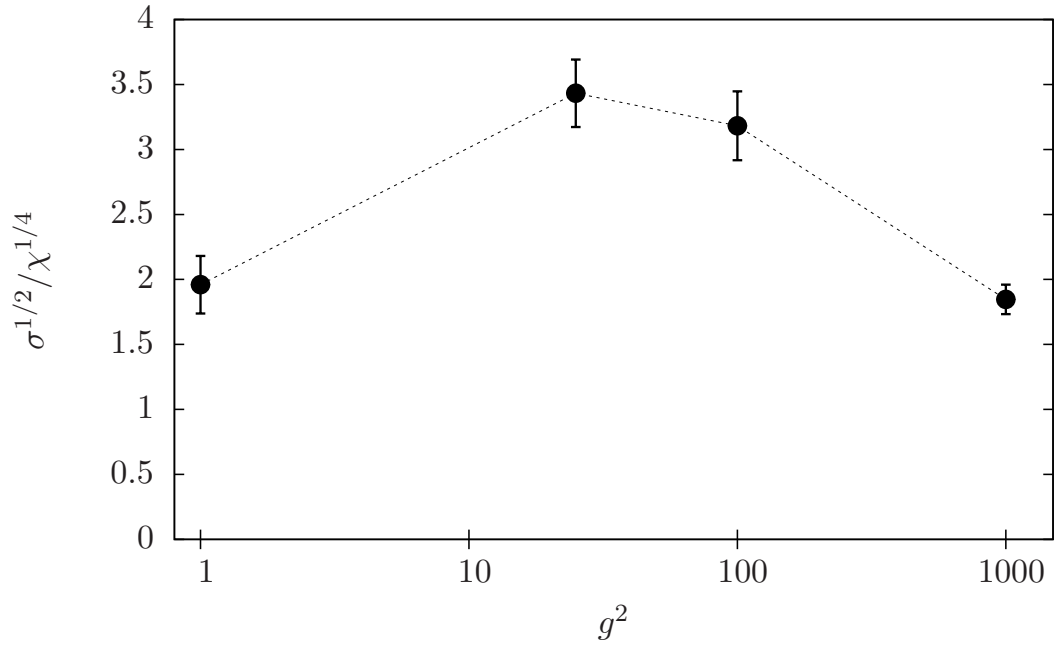}
\end{center}
\caption{The dimensionless ratio $\protect\sigma ^{1/2}/\protect\chi ^{1/4}$
as a function of the square coupling $g^{2}$.}
\label{sigoverts}
\end{figure}

The third dimensionless quantity which can be formed from our results is $%
\chi ^{1/4}/\sigma ^{1/2}$. In contrast to the ratios discussed above, it
has the additional benefit of not involving the rather unreliably known and
renormalization-scale dependent expectation value of the action density. Our 
$\chi ^{1/4}/\sigma ^{1/2}$ values for the five different $g^{2}$ are listed
in Table \ref{respu}. The inverse ratio $\sigma ^{1/2}/\chi ^{1/4}$ is
plotted for the four finite $g^{2}$ values in Fig. \ref{sigoverts}. For $%
g^{2}$ between $1$ and $1000$ our values for $\chi ^{1/4}/\sigma ^{1/2}$ lie
in the range from $0.3$ to $0.55$. Hence they are indeed rather compatible
with the SU$(2)$ lattice value $\chi ^{1/4}/\sigma ^{1/2}=0.483\pm 0.006$ 
\cite{luc01}. Moreover, for $g^{2}=25-100$ our results are practically
identical to the meron-ensemble value $\chi ^{1/4}/\sigma ^{1/2}\simeq 0.31$ 
\cite{len08} at $g^{2}=32$ (and somewhat smaller than the range $0.42\leq
\chi ^{1/4}/\sigma ^{1/2}\leq 0.48$ obtained from regular-gauge instanton
ensembles \cite{len08}).

Since correlations among individual observables may be hidden in the
dimensionless ratios discussed above, it is also important to directly
compare dimensionful results in physical units with those of other
approaches. For this purpose, we have considered two alternative
scale-setting procedures. The first was introduced in Sec. \ref{tssec} and
prescribes a standard value for the topological susceptibility in SU$\left(
2\right) $ Yang-Mills theory. This allows to rewrite our results for the
string tension in physical units, as given in the last column of Table \ref%
{tstab}. For $g^{2}=25-100$ they are roughly twice as large as those
obtained from Regge phenomenology (see below), while for $g^{2}=1000$ our
value is about 25\% smaller. Continuing this trend, the string tension
becomes unnaturally small in the strong-coupling limit, as a consequence of $%
\hat{\chi}$ growing too large due to boundary artefacts (cf. Sec. \ref{tssec}
and below).

In the following, we therefore adopt a more standard approach to scale
setting which fixes the string tension at its physical value $\sigma \simeq
4.2$ fm$^{-2}$ (estimated from the almost universal experimental Regge
slopes \cite{mil}). This amounts to writing our discretization scale in
physical units as $a=\hat{a}\left( \hat{\sigma}/\sigma \right) ^{1/2}$, with 
$\hat{a}=0.1$ and $\hat{\sigma}$ (cf. Table \ref{wltab}) given in numerical
units. The resulting values in Table \ref{respu} show that $a$ increases
with $g^{2}$. Hence the resolution of short-distance features in the vacuum
field population decreases moderately with the coupling. This tendency is
enhanced by the gradual dissociation of the well-localized, singular-gauge
dimerons into broader merons (cf. Sec. \ref{dissocsec}) which contain less
short-wavelength Fourier modes. As argued above, we expect our ensembles to
best reproduce observables in the confining Yang-Mills vacuum phase for $%
g^{2}=25-100$. In contrast, our most weakly coupled ensemble with $g^{2}=1$
underestimates the string tension while for $g^{2}\gg 100$ the results
increasingly suffer from boundary artefacts. In those ensembles, fixing the
string tension at its physical value will therefore not yield dimensionful
results of the expected magnitude. With this caveat in mind, we can now use $%
a$ to express other dimensionful quantities in physical units (cf. Table \ref%
{respu}).

%TCIMACRO{\TeXButton{B}{\begin{table}[htbp] \centering}}%
%BeginExpansion
\begin{table}[htbp] \centering%
%EndExpansion
\begin{tabular}{|c|c|c|c|c|c|}
\hline
$g^{2}$ & $\chi ^{1/4}/\sigma ^{1/2}$ & $a$ [fm] & $\rho $ [fm] & $n$ [fm$%
^{-4}$] & $\chi ^{1/4}$ [MeV] \\ \hline\hline
\multicolumn{1}{|l|}{$1$ \ \ \ \ \ \ } & $0.52\pm 0.057$ & $0.073\pm 0.004$
& $0.040\pm 0.002$ & $15.69\pm 3.44$ & $209.02\pm 23.04$ \\ \hline
\multicolumn{1}{|l|}{$25$} & $0.29\pm 0.022$ & $0.129\pm 0.002$ & $0.071\pm
0.001$ & $1.71\pm 0.099$ & $117.61\pm 8.90$ \\ \hline
\multicolumn{1}{|l|}{$100$} & $0.32\pm 0.012$ & $0.142\pm 0.002$ & $0.078\pm
0.001$ & $1.10\pm 0.062$ & $127.24\pm 5.034$ \\ \hline
\multicolumn{1}{|l|}{$1000$} & $0.54\pm 0.033$ & $0.219\pm 0.004$ & $%
0.121\pm 0.002$ & $0.19\pm 0.014$ & $218.66\pm 13.44$ \\ \hline
\multicolumn{1}{|l|}{$\infty $} & $2.46\pm 0.13$ & $0.205\pm 0.01$ & $%
0.113\pm 0.006$ & $0.25\pm 0.049$ & $991.28\pm 51.42$ \\ \hline
\end{tabular}
\caption{Results in physical units set by the string tension $\sigma=4.2 {\rm fm}^{-2}$.
\label{respu}} 
%TCIMACRO{\TeXButton{E}{\end{table}}}%
%BeginExpansion
\end{table}%
%EndExpansion

An important dimensionful quantity is the meron-center size regulator $\rho $
which furnishes the characteristic input scale of the dimeron ensembles (cf.
Sec. \ref{reg}). For $g^{2}=25-100$ and $\hat{\rho}=\allowbreak 0.55$ we
find $\rho \simeq 0.07-0.08\,\ $fm, i.e. about a quarter of the typical
instanton size $\bar{\rho}_{I}\sim 0.3$ fm in (singular-gauge)
instanton-liquid models \cite{sch98}. This indicates that our regularization
procedure deforms the classical dimeron solutions only moderately and should
not impede a potentially semiclassical behavior of their superpositions. The
average dissociation $2\left\langle \left\vert a\right\vert \right\rangle
\sim \left( 1-1.6\right) \rho $ of the dimerons at $g^{2}=25-100$ increases
the effective dimeron\ size, on the other hand, and brings it closer to
typical instanton sizes. Finally, it is useful to note that\ $\mu \sim \rho
^{-1}\simeq 2.5-2.8$ GeV approximately sets the renormalization scale of our
ensembles. (Recall that $\rho $ acts as an UV regulator e.g. for the action
density which diverges logarithmically when $\rho \rightarrow 0$.) These $%
\mu $ values are of the order of typical lattice scales, which may help to
explain why our results for the scale-dependent action density are
compatible with lattice values.

We now turn to the dimeron density of our ensembles. Recalling the
grid-point distribution from Secs. \ref{box} and \ref{statsec}, the
four-volume of our ensemble box is given by $V_{\text{ens}}=\left(
1.15\times 25\right) ^{3}\times \left( 1.15\times 40\right) a^{4}$. The
corresponding values of the dimeron density $n=\left( N_{D}+N_{\overline{D}%
}\right) /V_{\text{ens}}=487/V_{\text{ens}}$ in physical units are listed in
Table \ref{respu}. In the preferred coupling\ region $g^{2}=25-100$ they are
about 20--80\% larger than the typical instanton densities $n_{I+\bar{I}%
}\simeq 1$ fm$^{-4}$ of\ instanton liquid\ models \cite{sch98}. The physical
significance of this result will be discussed below. Moreover, it is
interesting to note that the density $n_{M+\bar{M}}\simeq 3.2$ fm$^{-4}$ of
merons in the $N_{M}=200$ ensemble of Ref. \cite{len08} is similar to the
density of our meron constituents, i.e. approximately twice as large as our
dimeron densities $n\sim 1.2-1.8$. This adds to our earlier evidence for an
increasingly dynamical role of the largely independent meron constituents
from the farthest dissociated dimerons.

It remains to put our results for the topological susceptibility in the last
column of Table \ref{respu} into a quantitative perspective. The dimeron
ensemble predictions for $\chi ^{1/4}$ in the coupling\ region $g^{2}=25-100$
are about 50\% smaller than the \textquotedblleft
physical\textquotedblright\ value $\chi ^{1/4}=195.03~$MeV (i.e. the SU$%
\left( 2\right) $ lattice value with the scale set by $\sigma =4.2$ fm$^{-2}$%
, cf. Sec. \ref{tssec}). Similarly small values $\chi ^{1/4}\simeq 118-132~$%
MeV were encountered in the meron ensembles of Ref. \cite{len08} while
regular-gauge instanton ensembles yield $\chi ^{1/4}\simeq 162-190~$MeV.
Hence our relatively small $\chi ^{1/4}$ values are consistent with previous
indications for the meron centers, which carry only half of the dimerons'
topological charge, to increasingly determine the topological-charge
fluctuations for $g^{2}\gtrsim 25$. (One should also keep in mind, however,
that we consider our values for $\hat{\chi}$ as lower bounds, cf. Sec. \ref%
{tssec}.) Another suggestive pattern in our data is the weak bare-coupling
dependence of the $\chi ^{1/4}$ values for $g^{2}=25-100$. As already
alluded to, this behavior may reflect the well-known temperature
independence of the topological susceptibility in the confined Yang-Mills
phase \cite{luc04} which in turn indicates that $\chi ^{1/4}$ is relatively
insensitive to temperature-induced variations of the coupling $g_{\text{YM}%
}^{2}\left( T\right) $ up to the critical temperature $T_{c}$. (For $T>T_{c}$%
, on the other hand, $\chi $ drops very rapidly.)

The unusually large value of $\chi ^{1/4}$ in our $g^{2}=\infty $ ensemble
was already noted above in the raw data (cf. Table \ref{tstab}). Of course
this result raises suspicion. In a random distribution of merons, into which
the dimerons dissociate for $g^{2}\rightarrow \infty $ as far as our limited
field configuration space and the boundary allows, one would expect the
topological susceptibility to be an incoherent sum of single-pseudoparticle
contributions and thus to have a value of the order of the pseudoparticle
density. This expectation was verified explicitly in random singular-gauge
instanton \cite{sch98}, regular-gauge instanton and single-meron \cite{len08}
ensembles \footnote{%
Even at finite coupling, correlations between instantons are known to be
fairly weak in pure Yang-Mills theory and in singular-gauge instanton
ensembles (without light quarks). Hence the topological susceptibility can
be used to estimate the instanton density in the Yang-Mills vacuum \cite%
{sch98}.}. The fact that our for $g^{2}\rightarrow \infty $ strongly
increasing $\chi $ values fail to scale with the density is therefore
probably another manifestation of the residual correlations between the
dimerons' meron partners and the related $g^{2}\rightarrow \infty $ boundary
artefacts. We recall that in Secs. \ref{dissocsec} and \ref{tssec} these
artefacts were found to impede the development of a genuine random meron
ensemble and to strongly contaminate the topology distributions.
Fortunately, in the physically relevant coupling region $g^{2}=25-100$ these
effects are much weaker and under far better control.

Our above results suggest a remarkably comprehensive role for dimerons in
the Yang-Mills vacuum. A key feature of the emerging picture is the rather
distinct division of labour between almost fully contracted, instanton-like
dimerons and their far dissociated, meron-like counterparts. In dynamical
equilibrium at intermediate, i.e. physical couplings both of these
components coexist at fairly high densities \footnote{%
Indications for the coexistence of instantons and merons at comparable
densities in the Yang-Mills vacuum were also found in a variational approach 
\cite{var}.}. In particular, our overall dimeron densities\ significantly
exceed the instanton density in instanton liquid models. Hence the
instanton-like dimeron component should be sufficiently populated to perform
essentially all established tasks of instantons in the Yang-Mills vacuum 
\cite{sch98,instlat}. The strongly quadratic rise of our heavy-quark
potentials at small interquark separations provides additional evidence in
this direction (cf. Sec. \ref{wlsec}). The meron-like component, on the
other hand, seems to be mainly responsible for the longest-distance vacuum
physics including, most importantly, linear quark confinement as established
in Sec. \ref{wlsec} and in single-meron ensembles.

On a structural level, the above reinterpretation of the instanton component
in the Yang-Mills vacuum as mainly consisting of hardly dissociated dimerons
is compatible with the existing instanton phenomenology as well.\ Indeed,
due to the destructive interference between their meron-center tails the
instanton-like dimerons experience much weaker long-range interactions than
the almost liberated merons of the meron-like component (cf. Secs. \ref%
{dimenssec} -- \ref{colcorsec}). Similar to singular-gauge instantons in
sufficiently dilute superpositions they overlap less with each other,
furthermore, and thus retain more of the classical solutions' shapes and
semiclassical behavior. Hence the instanton-like dimeron component indeed
resembles a rather weakly interacting \textquotedblleft
liquid\textquotedblright . On the other hand, this raises the question why
cooling studies, designed to detect semiclassical solutions in equilibrated
lattice configurations, have reported clear evidence for instantons or
calorons with non-trivial holonomy but none for dimerons \cite%
{cal-latt,instlat}. Closer inspection reveals, however, that these findings
are not in conflict with our above scenario. In fact, the finite-resolution
algorithms used to identify these solutions by their shapes may have counted
small dimerons as instantons. Moreover, under cooling the logarithmic
attraction between the dimerons' (smoothed) meron centers can no longer be
counterbalanced by their decreasing entropy. Hence the dimerons will shrink
and at least partially coalesce into instantons \cite{ste00} as which they
are identified afterwards.

As established in Sec. \ref{wlsec}, the effectively released meron centers
of the meron-like dimeron component develop the strong long-range
correlations required to generate confinement. Although these color
correlations produce a more complex long-distance structure than encountered
in fully disordered random ensembles, they still generate enough entropy for
a string tension of the physically expected magnitude to develop (cf. Sec. %
\ref{wlsec}). Moreover, it seems likely that our above scenario can be
extended to an at least qualitatively realistic description of the QCD
vacuum. The instanton-like component of dimeron ensembles, in particular,
should adapt to the presence of light quarks in essentially the same manner
as in instanton liquid models. This is because most instanton effects in the
light-quark sector, including the induced quark interactions which
spontaneously break chiral symmetry, originate from the characteristic quark
zero-modes arising in the instanton background. Since these zero modes are
of topological origin, their existence and bulk properties \cite{kis78} are
little affected by moderate deformations of the instantons into dimerons.
Hence the ensuing phenomenology should be similarly robust under such
deformations.

\section{Summary and conclusions}

\label{sum}

We have studied the extent to which the Yang-Mills vacuum can be described
by restricting its gauge-field content to\ superpositions of regularized\ SU$%
\left( 2\right) $\ (anti-) dimeron solutions. Our set-up and analysis of the
corresponding effective theory focused mainly on structural changes in
dimeron ensembles as a function of the bare gauge coupling and traced their
impact on topological and confinement properties. The underlying quantum
superpositions of dimerons, with their dynamics governed by the Yang-Mills
action, were generated without recourse to weak-coupling or low-density
approximations by Monte-Carlo simulations in a wide range of bare coupling
values. The localization of the individual dimerons was enhanced by
transforming them into a singular gauge, which simultaneously improves
numerical efficiency, the vacuum description at shorter distances and the
compatibility with semiclassically motivated instanton liquid models.

Our initial survey of dimeron ensemble properties and their coupling
dependence concentrated on the spacial distribution of the dimerons, their
local degree of selfduality and their meron substructure. To understand
relevant aspects of the underlying inter-dimeron dynamics, we further
studied short- and intermediate-range correlations between positions,
topological charges and color orientations of neighboring dimerons and of
their meron centers. The resulting structural patterns turned out to be
multifaceted and revealing. A strong repulsion up to distances of the meron
center sizes governs the short-range interactions between dimerons of both
topological charges and at all couplings. In our minimally coupled and thus
most ordered ensemble, we additionally find a striking (about sevenfold)
preference for nearest-neighbor dimerons to carry opposite topological
charges, indicating a strong intermediate-range attraction (repulsion)
between dimerons of opposite (equal) topological charge. A less pronounced
but still significant majority of next-to-nearest neighbors is found to
carry identical topological charges. The resulting short- to mid-range order
among the topological charges is reminiscent of Debye screening clouds in a
plasma. These ordering patterns can be traced to the dependence of the
underlying interactions on the relative color orientations between
neighboring dimerons. Ensemble and configuration averages of these color
orientations indicate, on the other hand, that in the mean the
intermediate-range attraction between neighboring dimerons and anti-dimerons
is almost completely compensated by their shorter- and longer-range
repulsion and by the ensemble entropy. In any case, with increasing coupling
and hence entropy\ the observed topological order weakens rather rapidly,
together with the underlying color correlations.

The most far-reaching impact of the increasing bare coupling, however,
manifests itself in the structural changes which it induces in the
individual dimerons. In particular, we found the first robust evidence,
established in fully interacting ensembles, for the dimerons to gradually\
dissociate into their meron constituents when the coupling grows. This
process is driven by a competition between the coupling-dependent
interactions and the entropy. Several complementary pieces of evidence
further indicate that\ the meron constituents continuously replace the
dimerons as the dynamically relevant degrees of freedom. With growing
coupling\textbf{\ }the interactions between merons belonging to different
dimerons therefore increasingly determine the ensemble properties. This
renders the dimerons' meron partners more and more independent (except for
their frozen relative color orientation) and allows our field configurations
to approximate more closely those of meron ensembles. In particular,
however, the changing nature of the dynamically active topological-charge
carriers and of their coupling-dependent correlations leaves various
revealing imprints on bulk ensemble properties.

Among the latter, the\ topological susceptibility is particularly relevant
in our context. When calculated in a sequence of growing evaluation volumes
up to the largest accessible ones, it is found to level off towards the end
of the sequence. This indicates or at least foreshadows the expected
saturation of the topological-charge fluctuations in the thermodynamic
limit. Moreover, the resulting values of the susceptibility in the
intermediate, physical coupling range remain within errors practically
constant. This behavior may be related to the similarly weak temperature
dependence of the\ topological susceptibility in Yang-Mills theory below the
deconfinement temperature. Our intermediate-coupling results reproduce those
of comparable single-meron ensembles rather closely, furthermore, which
seems to indicate that (on average relatively loosely\ bound) meron centers
of dimerons and single merons produce topological charge fluctuations of
roughly equal strength. On the other hand, our results for the topological
susceptibility underestimate lattice and instanton ensemble values by about
50\%, perhaps due to insufficient saturation in our evaluation volumes and
the underlying boundary effects. In any case, the dimeron ensemble
predictions for the susceptibility could be straightforwardly adjusted by
increasing the dimeron density.

Since the long-range color tails of less bound meron centers disorder the
vacuum more strongly, we have payed particular attention to the confinement
properties of the dimeron ensembles. The latter were monitored by evaluating
ensemble averages of rectangular Wilson loops and by extracting the
associated static quark-antiquark potentials for a wide range of coupling
values. Initially, the resulting potentials are found to rise quadratically
with the interquark separation. This behavior is typical for (sufficiently
dilute) instanton ensembles and indicates that our short-distance potentials
are dominated by contributions from moderately dissociated dimerons which
act essentially as singular-gauge instantons. For quark-antiquark
separations beyond about half a fermi, on the other hand, the Wilson loops
indeed develop an area-law behavior and the potentials consequently show the
linear distance dependence characteristic of quark confinement. The
associated string tension is found to grow with the coupling and to reach
values of the experimentally expected order of the magnitude in the physical
coupling range.

Hence we have shown -- without reliance on weak-coupling or low-density
approximations -- that sufficiently strongly coupled and therefore on
average far enough dissociated dimeron ensembles indeed produce confining
long-range correlations of about the physically required strength. Since the
latter originate from the meron tails of the more strongly dissociated
dimerons, these results also provide complementary evidence for single-meron
ensembles to confine (even though our finite simulation volume and our
limited field configuration space prevent the dimerons from breaking up
completely). Moreover, both dimeron and single-meron ensembles are found to
generate confining static-quark potentials of comparable strengths at our
largest accessible interquark distances. Since the linear rise of the
meron-induced potential turns out to continue beyond these distances,
furthermore, our above reasoning indicates that the same should happen in
dimeron ensembles. Hence we expect an ongoing linear rise of the
dimeron-induced potentials into the distance region where the underlying
color flux tubes are fully developed.

While dimeron ensembles thus share with single-meron ensembles the main
features of the\ confinement mechanism, they additionally provide a more
complete description of the Yang-Mills vacuum. The first evidence is that
our results establish how a confining meron component in the vacuum can
robustly emerge from instanton and dimeron dissociation. This provides, in
particular, the first quantitative confirmation for a key ingredient of the
meron-induced confinement scenario envisioned more than three decades ago on
the basis of qualitative arguments. In addition, our results on the coupling
dependence of dimeron ensemble properties should contain useful information
on their temperature dependence and, in particular, on the impact of
temperature-induced variations of the gauge coupling (although we did not
pursue this issue in the present paper).

Furthermore, we found evidence for the physical dimeron density to be about
twice as large as typical instanton densities. This allows confining dimeron
ensembles to retain a phenomenologically relevant fraction of contracted and
thus more weakly interacting, \textquotedblleft
instanton-like\textquotedblright\ dimerons. An independent indication for
their considerable density is the strong quadratic rise of our heavy-quark
potentials at short interquark distances. The instanton-like component
should adapt to the presence of light quarks in essentially the same fashion
as in phenomenologically successful instanton vacuum models, furthermore,
since the underlying quark zero-modes in the instanton background are of
topological origin and therefore little affected by the moderate
deformations of instantons into contracted dimerons. Hence, in addition to
confining color, dimeron ensembles should be able to reproduce much of the
successful vacuum and hadron phenomenology predicted by (non-confining)
singular-gauge instanton ensembles.

\begin{acknowledgments}
We would like to thank E.-M. Ilgenfritz and M. Wagner for their help and
numerous intense discussions, and we are grateful to M. Wagner for providing
us with his pseudoparticle simulation code. H.F. and M.M.-P. acknowledge
financial support from the Deutsche Forschungsgemeinschaft (DFG) under
project Mu932/6-1.
\end{acknowledgments}

\newpage

\begin{figure}[tbp]
\begin{center}
\includegraphics[height = 8cm]{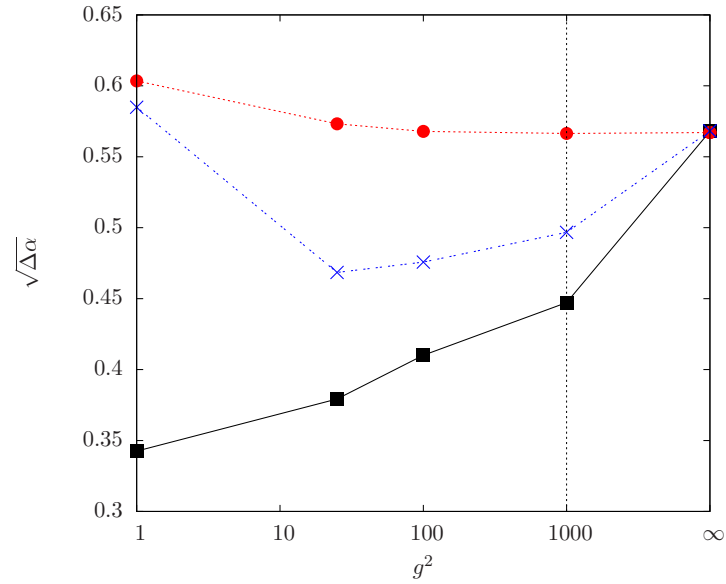}
\end{center}
\caption{The standard deviation of the $\left\langle \protect\alpha %
\right\rangle $ distribution between nearest neighbors of equal (black
squares, full line) and opposite (red bullets, dotted line) topological
charge $Q$ as a function of $g^{2}$. The standard deviation of the color
orientation for nearest neighbors independent of $Q$ is also included (blue
crosses, dashed line).}
\label{stderfig}
\end{figure}

\bigskip

\end{document}